\documentclass[a4paper,envcountsame]{llncs}

\usepackage{mypackages}
\usepackage{mycommands}
\title{Security Analysis of the EDHOC protocol}
\author{Baptiste Cottier \and David Pointcheval}
\institute{DIENS, {\'E}cole normale sup{\'e}rieure, CNRS, Inria, PSL University, Paris, France}
\date{}
\begin{document}

\maketitle

\begin{abstract}
    Ephemeral Diffie-Hellman Over COSE (\EDHOC) aims at being a very compact and lightweight authenticated Diffie-Hellman key exchange with ephemeral keys.
    It is expected to provide mutual authentication, forward secrecy, and identity protection, with a 128-bit security level.
    
    A formal analysis has already been proposed at SECRYPT '21, on a former version, leading to some improvements, in the ongoing evaluation process by IETF. Unfortunately, while formal analysis can detect some misconceptions in the protocol, it cannot evaluate the actual security level.
    
    In this paper, we study the last version. Without complete breaks, we anyway exhibit attacks in $2^{64}$ operations, which contradict the expected 128-bit security level. We thereafter propose improvements, some of them being at no additional cost, to achieve 128-bit security for all the security properties (i.e. key privacy, mutual authentication, and identity-protection).
\end{abstract}

\section{Protocol Description}
\label{sec:description}

Ephemeral Diffie-Hellman over COSE \cite{COSE} (\EDHOC) aims to provide a common session key to two parties potentially running on constrained devices over low-power IoT radio communication technologies. 
\EDHOC protocol can be instantiated with several settings: 
\begin{itemize}
    \item \emph{Authentication Method}: Each party (Initiator and Responder) can use an authentication method: either with a signature scheme (\SIG), or with a static Diffie-Hellman key (\STAT).
    \begin{center}
        \begin{tabular}{|c|c|c|} \hline
            Value & Initiator & Responder \\ \hline
            0 & \SIG: Signature & \SIG: Signature \\
            1 & \SIG: Signature & \STAT: Static DH \\
            2 & \STAT: Static DH & \SIG: Signature \\
            3 & \STAT: Static DH & \STAT: Static DH \\ \hline
        \end{tabular}
    \end{center}  
    \item \emph{Cipher Suites}: Ordered set of protocol security settings. Initial paper offers many possible suites, but we focus on the most aggressive cipher suites setting the MAC length to 8 bytes, i.e. Cipher Suites 0 and 2 which share the following parameters: 
    \begin{center}
        \begin{tabular}{|c|c|c|c|} \hline
            (Application) AEAD & Hash & MAC len\\ \hline
            AES-CCM-16-64-128 & SHA-256& 8 \\ \hline
        \end{tabular}
    \end{center}  
    The difference between Cipher Suites 0 and 2 is the Elliptic Curve used: X25519 in suite 0 and P-256 in suite 2. 
    \item \emph{Connection Identifiers.} Data that may be used to correlate between messages and facilitate retrieval of protocol state in \EDHOC and application.
    \item \emph{Credentials and Identifiers.} They are used to identify and optionally transport the authentication keys of the Initiator and the Responder.
\end{itemize}
We suppose both the Initiator and the Responder are aware that the authentication method is 3, the \STAT/\STAT. Also, as said before, the difference between Cipher Suite 0 and Cipher Suite 2 is the choice of the Elliptic Curve. As both curves provide the same security guarantee and are more an implementation concern, we do not include the Cipher Suite ID \texttt{Suites\_I} (as it appears in \cite{EDHOC-draft}) in the first message of the protocol. 

\paragraph{Extract and Expand.} 
In the \EDHOC Key-Schedule defined below, the pseudorandom keys (\PRK) are derived using an extraction function.
In our context, as we study cipher suites 0 and 2, our hash algorithm is set as SHA-256.
Therefore $\sfExt(\salt,\IKM)=\HKDFExt(\salt,\IKM)$ is defined with SHA-256, where $\IKM$ holds for Input Keying Material (in our context, this will be some Diffie-Hellman keys) and $\sfExp(\PRK, \info, \len) = \HKDFExp(\PRK, \info, \len)$ where $\info$ contains the transcript hash ($\TH_2$, $\TH_3$ or $\TH_4$), the name of the derived key and some context, while $\len$ denotes the output length.

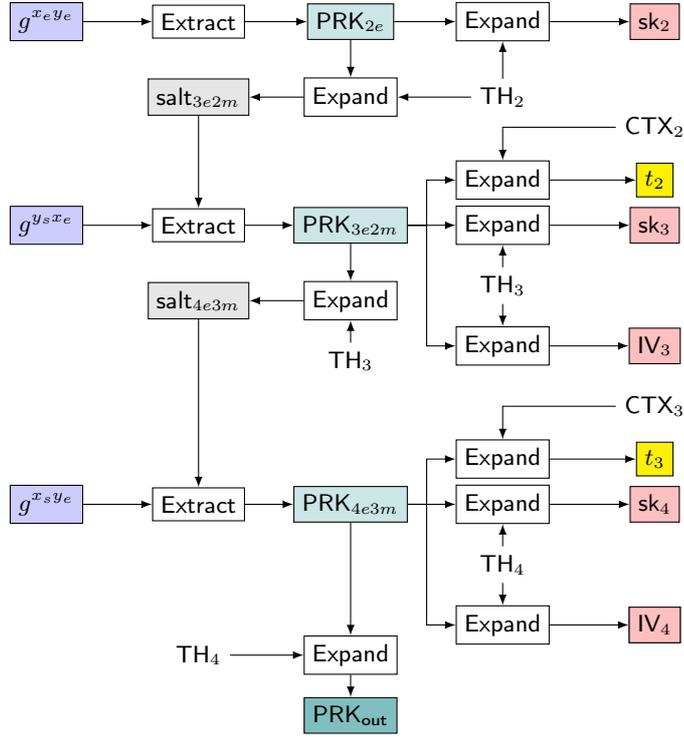
\begin{figure}[ht] \centering
    \begin{tikzpicture}
        \definecolor{kellygreen}{rgb}{0.3, 0.73, 0.09}
         
        \node[draw, fill=blue!20]   at  (0  ,8.2)   (GXY)       {$g^{x_e y_e}$}         ; 
        \node[draw, fill=blue!20]   at  (0  ,5.5)   (GRX)       {$g^{y_s x_e}$}         ; 
        \node[draw, fill=blue!20]   at  (0  ,1.8  )   (GIY)       {$g^{x_s y_e}$}         ;

        \node[draw]                 at  (2  ,8.2)   (XT2)       {\sfExt}                ;
        \node[draw]                 at  (2  ,5.5)   (XT3)       {\sfExt}                ;
        \node[draw]                 at  (2  ,1.8  )   (XT4)       {\sfExt}                ;

        \node[draw, fill=teal!20]   at  (4  ,8.2)   (PRK2)      {$\PRK_{2e}$}           ; 
        \node[draw, fill=teal!20]   at  (4  ,5.5)   (PRK3)      {$\PRK_{3e2m}$}         ; 
        \node[draw, fill=teal!20]   at  (4  ,1.8  )   (PRK4)      {$\PRK_{4e3m}$}         ;
        \node[draw, fill=teal!50]  at  (4  ,-1  )   (PRKout)    {$\PRK_{\out}$}         ;
        
        \node[draw]                 at  (6  ,8.2)   (XP2)       {\sfExp}                ;
        \node[draw]                 at  (6  ,6.1)   (XP3t)       {\sfExp}                ;
        \node[draw]                 at  (6  ,5.5)   (XP3sk)      {\sfExp}                ;
        \node[draw, fill=white]                 at  (6  ,3.9)   (XP3IV3)      {\sfExp}                ;
        \node[draw]                 at  (6  ,2.4)   (XP4t)      {\sfExp}                ;
        \node[draw]                 at  (6  ,1.8)   (XP4sk4)       {\sfExp}                ;
        \node[draw]                 at  (6  ,0.2)   (XP4IV4)       {\sfExp}                ;
        \node[draw]                 at  (4  ,7.2)   (XPs3)      {\sfExp}                ;
        \node[draw]                 at  (4  ,4.5)   (XPs4)       {\sfExp}                ;
        \node[draw]                 at  (4  ,-0.2)   (XPout)      {\sfExp}                ;
        
        \node[draw, fill=gray!20]   at  (2  ,7.2)   (s3e2m)     {$\salt_{3e2m}$}        ;
        \node[draw, fill=gray!20]   at  (2  ,4.5)   (s4e3m)     {$\salt_{4e3m}$}        ;

        \node                       at  (6  ,7.2)   (TH2)       {$\TH_2$}               ;
        \node                       at  (6  ,4.7)   (TH3)       {$\TH_3$}               ;
        \node                       at  (4  ,3.7)   (TH3b)       {$\TH_3$}               ;
        \node                       at  (6  ,1)   (TH4)       {$\TH_4$}               ;
        \node                       at  (2  ,-0.2)   (TH4b)       {$\TH_4$}               ;

        \node                       at  (8  ,6.8)   (ctx2)      {$\textsf{CTX}_2$}  ;
        \node                       at  (8 ,3.1)   (ctx3)      {$\textsf{CTX}_3$}  ;

        \node[draw, fill=pink]      at  (8  ,8.2)   (sk2)       {$\sk_2$}               ;        
        \node[draw, fill=yellow]    at  (8  ,6.1)   (t2)        {$t_2$}                 ;
        \node[draw, fill=pink]      at  (8  ,5.5)   (sk31)       {$\sk_3$}      ;
        \node[draw, fill=pink]      at  (8  ,3.9)   (IV3)       {$\IV_3$}      ;
        \node[draw, fill=pink]      at  (8  ,0.2)   (IV4)       {$\IV_4$}      ;
        \node[draw, fill=yellow]    at  (8  , 2.4)   (t3)        {$t_3$}                 ;
        \node[draw, fill=pink]      at  (8  ,1.8)   (sk4)       {$\sk_4$}      ;
        \node[draw, fill=pink]      at  (8  ,0.2)   (IV4)       {$\IV_4$}      ;
        
        \draw[-latex] (GXY.east) -- (XT2.west) ; 
        \draw[-latex] (XT2.east)  -- (PRK2.west); 
        \draw[-latex] (PRK2.east) -- (XP2.west) ; 
        \draw[-latex] (XP2.east) -- (sk2.west) ; 
        \draw[-latex] (PRK2.south) -- (XPs3.north) ;
        \draw[-latex] (XPs3.west) -- (s3e2m.east) ; 
        
        \draw[-latex] (s3e2m.south) -- (XT3.north) ;

        \draw[-latex] (GRX.east) -- (XT3.west) ; 
        \draw[-latex] (XT3.east)  -- (PRK3.west); 
        \draw[-latex] (PRK3.east) -- (5,5.5) -- (5,6.1) -- (XP3t.west) ; 
        \draw[-latex] (PRK3.east) -- (XP3sk.west) ; 
        \draw[-latex] (PRK3.east)  -- (5,5.5) -- (5,3.9) --  (XP3IV3.west) ; 
        \draw[-latex] (XP3t.east) -- (t2.west) ; 
        \draw[-latex] (XP3sk.east) -- (sk31.west) ; 
        \draw[-latex] (PRK3.south) -- (XPs4.north) ;
        \draw[-latex] (XPs4.west) -- (s4e3m.east) ; 

        \draw[-latex] (s4e3m.south) -- (XT4.north) ;

        \draw[-latex] (GIY.east) -- (XT4.west) ; 
        \draw[-latex] (XT4.east)  -- (PRK4.west); 
        \draw[-latex] (PRK4.east) -- (5, 1.8) -- (5,2.4) -- (XP4t.west) ; 
        \draw[-latex] (PRK4.east) --  (XP4sk4.west) ; 
        \draw[-latex] (PRK4.east)  -- (5,1.8) -- (5,0.2) --  (XP4IV4.west) ; 
        \draw[-latex] (XP4t.east) -- (t3.west) ; 
        \draw[-latex] (XP4sk4.east) -- (sk4.west) ; 
        \draw[-latex] (XP4IV4.east) -- (IV4.west) ; 
        \draw[-latex] (XP3IV3.east) -- (IV3.west) ; 
        \draw[-latex] (PRK4.south) -- (XPout.north) ;
        \draw[-latex] (XPout.south) -- (PRKout.north) ;
        \draw[-latex] (TH2.north) -- (XP2.south) ; 
        \draw[-latex] (TH2.west) -- (XPs3.east) ; 

        \draw[-latex] (TH3.north) -- (XP3sk.south) ; 
        \draw[-latex] (TH3.south) -- (XP3IV3.north) ; 
        \draw[-latex] (TH3b.north) -- (XPs4.south) ; 

        \draw[-latex] (TH4.north) -- (XP4sk4.south) ; 
        \draw[-latex] (TH4.south) -- (XP4IV4.north) ; 
        \draw[-latex] (TH4b.east) -- (XPout.west);
        \draw[-latex] (ctx2.west) -- (6,6.8) -- (XP3t.north) ; 
        \draw[-latex] (ctx3.west) -- (6,3.1) -- (XP4t.north) ; 
        
    \end{tikzpicture}
    \caption{Key Derivation (for the \STAT-\STAT Method) from~\cite{EDHOCFormal21}. \label{Fig:Key-Schedule}}
\end{figure}

\paragraph{Key Schedule.} 
During the key exchange, several cryptographic computations occur. The keying material, including MAC Keys (as we study \STAT/\STAT method), Encryption Keys and Initialisation Vectors result from a key schedule, adapted from Norrman et al. \cite{EDHOCFormal21}.
In Fig. \ref{Fig:Key-Schedule}, purple boxes denote Diffie-Hellman shared secrets ($g^{x_e y_e},g^{y_s x_e},g^{x_s y_e}$), where $x_e$ and $y_e$ denote the ephemeral DH keys, while $x_s$ and $y_s$ are the long-term static DH keys. From those Diffie-Hellman shared secrets, we extract the pseudo random keys $\PRK_{2e}, \PRK_{3e2m}$ and $\PRK_{4e3m}$, in light blue. Those keys are then expanded to compute the encryption material (keys and \IV), in red, and the authentication tags $t_2$ and $t_3$, in yellow. Keys $\PRK_{2e}$ and $\PRK_{3e2m}$ are also used to compute the salts $\salt_{3e2m}$ and $\salt_{4e3m}$ respectively, in grey, used in the \HKDFExt function. The final session key $\PRK_{\out}$, backgrounded in green, is computed calling \HKDFExp on $\PRK_{4e3m}$.
Transcript hashes, denoted $\TH_i$, are used as input to the \HKDFExp function. More precisely, with SHA-256 as $\cH$, we have:
\begin{align*}
    \TH_2 & =\cH(Y_e, \CR, \cH(m_1)) & \TH_3 & = \cH(\TH_2 , m_2) & \TH_4 & = \cH(\TH_3, m_3)
\end{align*}
where $m_1$ is the first message sent by the Initiator, $m_2$ and $m_3$ respectively are the plaintexts respectively encrypted in the message 2 and message 3.

\paragraph{Authenticated Encryption with Associated Data (AEAD).}
As said above, the cipher suites we work on both use AES-CCM-16-64-128. We detail this Authenticated Encryption scheme:
\begin{itemize}
    \item AES-CCM: CCM, for Counter with CBC-MAC is an AES mode providing both encryption and authentication. CCM mode combines the CBC-MAC and the CTR (counter) mode of encryption. The first step consists in calculating a tag $T$, then, encrypt the message and the tag using the counter mode. 
    \item 16: messages length is limited to $2^{16}$ bytes long (64KiB). Therefore, the nonce is 13 bytes long allowing $2^{13 \times 8}$ possibles values of the nonce without repeating
    \item 64: Tag is 64 bits long. 
    \item 128: Key is 128 bits long.
\end{itemize}  

\paragraph{Connection Identifiers (from \cite{EDHOC-draft}).} Connection identifiers ($\CI$ and $\CR$) may be used to correlate \EDHOC messages and facilitate the retrieval of protocol state during \EDHOC protocol execution or in a subsequent application protocol. The connection identifiers do not have any cryptographic purpose in \EDHOC.


\paragraph{\EDHOC-Exporter and \EDHOC-KeyUpdate.} At the end of the protocol, the Initiator and the Responder compute $\TH_4$. This value can be used in case an application need to export the \EDHOC session key. Also,  in case the key needs to be updated, the Initiator and the Responder rerun \HKDFExt, with $\PRK_{4e3m}$ as input, together with a random nonce agreed upon by the Initiator and the Responder. 

\paragraph{Protocol.} The detailed description of the protocol is given in Fig~\ref{Fig:EDHOC-description}. The final session key is $\SK = \PRK_{\out}$

\begin{figure}[htb]
    \begin{tabular}[h!]{|c|l|} \hline
        $\bG=<g>$ & Cyclic group generated by $g$ \\ \hline 
        $p$ & Size of the group $\bG$ \\ \hline 
        $\cH$ & Hash function SHA-256 (256 bits digest)\\ \hline
        $X_e, x_e$ & Initiator Ephemeral DH Public and Secret Key \\ \hline
        $X_s, x_s$ & Initiator Static DH Public and Secret Key \\ \hline
        $Y_e, y_e$ & Responder Ephemeral DH Public and Secret Key \\ \hline
        $Y_s, y_s$ & Responder Static DH Public and Secret Key \\ \hline
        $\EAD$ & External Authorized Data \\ \hline
        $ \sk $ & Secret key \\ \hline 
        $\cE , \cD $ & One-time Encryption and Decryption \\ \hline 
        $\cE' , \cD' $ & Authenticated Encryption and Decryption with Associated Data\\ \hline 
        $ \bot $ & Protocol abortion \\ \hline 
        $\TH$ & Transcript Hash \\ \hline
        $t_2, t_3$ & MAC tags \\ \hline
        $\CR, \CI$ & Connection Identifiers \\ \hline
        $ \lMAC$ & MAC output length \\ \hline
        $ \lPT$ & Plaintext length of the first message send by the responder \\ \hline
        $ \lH$ & Hash length \\ \hline
        $ \lkey, \lIV$ & Key and IV length \\ \hline
    \end{tabular}
    \caption{Notations}
\end{figure}

\begin{figure}
    \resizebox{\textwidth}{!}{
    \small
    \begin{tabular}{lcl}
       \multicolumn{1}{c}{Initiator} & & \multicolumn{1}{c}{Responder} \\ \hline 
       \multicolumn{1}{c}{$X_s = g^{x_s}$} & static keys & \multicolumn{1}{c}{$Y_s = g^{y_s}$} \\
        && \\
        $\underline{\initrun1(\ID_\sfI )}$ && \\
        $x_e \getsr \bZ_p, X_e \gets g^{x_e}$ \\
        $  \sfC_\sfI \getsr \bit^{nl} $ \\
        $m_1 \gets (X_e \| \CI \| \EAD_1)$ 

        & \sendright{$m_1$}{15mm}  & $\underline{\resprun1(\ID_\sfR, y_s, m_1)}$ \\

        &&  Parse $m_1$ as $(X_e \| c \| \EAD_1)$  \\
        && $y_e \getsr \bZ_p, Y_e \gets g^{y_e}$ \\
        && $ \CR \getsr \bit^{nl} $ \\
        && $ \sid \gets (\CI , \CR , X_e , Y_e)$ \\
        && $\PRK_{2e} \gets \HKDFExt( "" , {X_e}^{y_e})$ \\
        && $\TH_2 \gets \cH(Y_e, \CR, \cH(m_1))$ \\ 
        && $\sk_2 \gets \HKDFExp(\PRK_{2e}, 0, \TH_2, \lPT)$ \\
        && $\salt_{3e2m} \gets \HKDFExp(\PRK_{2e}, 1, \TH_2, \lH)$ \\
        && $\PRK_{3e2m} \gets \HKDFExt(\salt_{3e2m}, {X_e}^{y_s})$ \\
        && $\CTX_2 \gets (\ID_\sfR \| \TH_2 \| Y_s \| \EAD_2)$ \\
        && $t_2 \gets \HKDFExp(\PRK_{3e2m} , 2, \CTX_2, \lMAC) $ \\
        && $m_2 \gets (\ID_\sfR \| t_2 \| \EAD_2 )$ \\
        
        $\underline{\initrun2(\ID_\sfI, x_s, Y_s,(Y_e, c_2, \CR) )}$ & \sendleft{$Y_e , c_2, \CR$}{15mm} & $c_2 \gets\cE(\sk_2 , m_2)$ \\
        
        $\PRK_{2e} \gets \HKDFExt( "" , {Y_e}^{x_e})$ \\
        $\TH_2 \gets \cH(Y_e, \CR, \cH(m_1))$ \\
        $ \sk_2 \gets \HKDFExp(\PRK_{2e}, 0, \TH_2, \lPT)$ \\
        Set $m_2 \gets \cD(\sk_2 , c_2)$ \\
        Parse $m_2$ as $(\ID_\sfR \| t_2 \| \EAD_2)$ \\       
        $\CTX_2 \gets (\ID_\sfR \| \TH_2 \| Y_s \| \EAD_2)$ \\
        $\salt_{3e2m} \gets \HKDFExp(\PRK_{2e}, 1, \TH_2, \lH)$ \\
        $\PRK_{3e2m} \gets \HKDFExt(\salt_{3e2m}, {Y_s}^{x_e})$ \\
        $t'_2 \gets \HKDFExp(\PRK_{3e2m} , 2, \CTX_2, \lMAC) $ \\
        if $t'_2 \neq t_2$ : \textbf{return} $\bot$ \\ 
        $\TH_3 \gets \cH(\TH_2 , m_2)$\\
        $\sk_3 \gets \HKDFExp(\PRK_{3e2m}, 3, \TH_3, \lkey)$ \\ 
        $\IV_3 \gets \HKDFExp(\PRK_{3e2m}, 4,  \TH_3, \lIV)$ \\
        $\salt_{4e3m} \gets \HKDFExp(\PRK_{3e2m}, 5, \TH_3, \lH)$ \\
        $\PRK_{4e3m} \gets \HKDFExt(\salt_{4e3m}, {Y_e}^{x_s})$ \\
        $\accepted \gets 1$ \\
        $\CTX_3 \gets (\ID_\sfI \| \TH_3 \| X_s \| \EAD_3)$ \\
        $t_3 \gets \HKDFExp(\PRK_{4e3m}, 6, \CTX_3, \lMAC)$ \\
        $m_3 \gets (\ID_I \| t_3 \| \EAD_3)$ \\
        $c_3 \gets \cE'(\sk_3, \IV_3, m_3)$ & \sendright{$c_3$}{15mm} &  $\underline{\resprun2(\ID, \st, \peerpk, c_3)}$  \\
        
        && $\TH_3 \gets \cH(\TH_2 , m_2)$\\
        && $\sk_3 \gets \HKDFExp(\PRK_{3e2m}, 3, \TH_3, \lkey)$ \\ 
        && $\IV_3 \gets \HKDFExp(\PRK_{3e2m}, 4,  \TH_3, \lIV)$ \\
        && Set $m_3 \gets \cD'(\sk_3 , \IV_2, c_3)$ \\
        && $\pcif m_3 = \bot : \pcreturn \bot $\\
        && Parse $m_3$ as $(\ID_\sfI \| t_3 \| \EAD_3)$ \\
        && $X_s \gets \peerpk[\ID_\sfI]$ \\
        && $\salt_{4e3m} \gets \HKDFExp(\PRK_{3e2m}, 5, \TH_3, \lH)$ \\
        && $\PRK_{4e3m} \gets \HKDFExt(\salt_{4e3m}, {X_s}^{y_e})$ \\
        && $\accepted \gets 1$ \\
        && $\CTX_3 \gets (\ID_\sfI \| \TH_3 \| X_s \| \EAD_3)$ \\
        &&  $t'_3 \gets \HKDFExp(\PRK_{4e3m}, 6, \CTX_3, \lMAC)$ \\
        && if $t'_3 \neq t_3$ : \textbf{return} $\bot$ \\
 
        $\TH_4 \gets \cH(\TH_3, m_3) $ && $\TH_4 \gets \cH(\TH_3, m_3) $ \\
        $\PRK_\out \gets \HKDFExp(\PRK_{4e3m}, 7, \TH_4, \lH)$ && $\PRK_\out \gets \HKDFExp(\PRK_{4e3m}, 7, \TH_4, \lH)$ \\
        $\terminated \gets 1$ &&  $\terminated \gets 1$ \\
        $\SK \gets \PRK_\out$ && $\SK \gets \PRK_\out$ \\
    \end{tabular}
    }
    \caption{\EDHOC (draft-ietf-lake-edhoc-15) in the \STAT/\STAT Authentication Method} \label{Fig:EDHOC-description}
\end{figure}

\section{Security Concerns}
\label{sec:security_concerns}
\paragraph{Security Goals.} The security goals of an authenticated key exchange protocol are:
\begin{itemize}
    \item \emph{Key Privacy}: Equivalent to Implicit Authentication. At most both participants know the final session key, which should remain indistinguishable from random to outsiders. With additional \emph{Perfect Forward Secrecy}, by compromising the long-term credential of either peer, an attacker shall not be able to distinghuish past session keys from random keys. In our context, this will rely on a Diffie-Hellman assumption.
    \item \emph{Mutual Authentication}: Equivalent to explicit authentication.  Exactly both participants have the material to compute the final session key.
    \item \emph{Identity Protection}: At most both participants know the identity of the Initiator and the Responder.
\end{itemize}

\paragraph{Random Oracle Model.} For the security analysis, we model Hash and Key Derivation Functions as random oracles. Respectively, the random oracles $\ROT$ and $\ROP$ will model $\HKDFExt$ and $\HKDFExp$ functions as perfect random functions.

\paragraph{Computational Diffie-Hellman Assumption (\CDH).}
The \CDH assumption in a group $\bG$ states that given $g^u$ and $g^v$, where $u$, $v$ were drawn at random from $\bZ_p$, it is hard to compute $g^{uv}$. 
This can be defined more precisely by considering an Experiment $\bfexp_\bG^{\CDH}(\cA)$, in which we select two values $u$ and $v$ in $\bZ_p$, compute $U=g^u$ and $V=g^v$, and then give both $U$ and $V$ to $\cA$. Let $Z$ be the output of $\cA$. Then, the Experiment $\bfexp_\bG^{\CDH}(\cA)$ outputs $1$ if $Z = g^{uv}$ and 0 otherwise. 
We define the advantage of $\cA$ in violating the \CDH assumption as $\Adv_\bG^{\CDH}(\cA) = \Pr[\bfexp_\bG^{\CDH}(\cA) = 1 ]$
and the advantage function of the group, $\Adv_\bG^{\CDH}(t)$, as the maximum value of $\Adv_\bG^{\CDH}(\cA)$ over all $\cA$ with time-complexity at most $t$.

\paragraph{Gap Diffie-Hellman (\GDH).} The Gap Diffie-Hellman problem aims to solve a \CDH instance $(g, U=g^u, V=g^v)$, as above, with access to a Decisional Diffie-Hellman oracle $\DDH$ returning 1 if a tuple $(g,g^a, g^b, g^c)$ is a Diffie-Hellman tuple, and 0 otherwise. 
We define the advantage function of the group $\Adv_\bG^{\GDH}(t, q_\DDH)$, as the maximum value of $\Adv_\bG^{\CDH}(\cA)$ over all $\cA$ with time-complexity at most $t$ and making at most $q_\DDH$ queries to the \DDH oracle. 
\paragraph{Symmetric Encryption.} In the following, we will use several symmetric encryption schemes, such as $\varPi=(\cE, \cD)$ with keys in $\cK$ and messages in $\cM$, with various properties:

\header{Injectivity.} $\varPi=(\cE, \cD)$ is injective if $$\forall k \in \cK, \cE(k,m_1)=\cE(k,m_2) \Longrightarrow m_1 = m_2.$$

\header{Semantic Security.}
$\varPi=(\cE, \cD)$ is semantically secure if, for chosen messages $m_0, m_1 \in \cM$, an adversary cannot distinguish $\cE(k, m_0)$ and $\cE(k, m_1)$ with a negligible advantage for a random key $k \in \cK$. This can be defined more precisely by considering the Experiment $\bfexp^{\ind}_{\varPi}(\cA)$, for indistinguishability, in which $\cA$ selects and gives us two messages $m_0$ and $m_1$, then we choose $b \in \bit$ and $k \in \cK$, compute and send $c=\cE(k,m_b)$ to $\cA$. Let $b' \in \bit$  be the output of $\cA$. Then, the Experiment $\bfexp^{\ind}_{\varPi}(\cA)$ outputs 1 if $b'=b$ and 0 otherwise. We define the advantage of $\cA$ in violating the semantic security of $\varPi$ as  $\Adv^{\ind}_{\varPi}(\cA) = \Pr[\bfexp^{\ind}_{\varPi}(\cA) = 1 ]$ and the advantage function $\Adv_\varPi^{\ind}(t)$, as the maximum value of $\Adv^{\ind}_{\varPi}(\cA)$ over all $\cA$ with time-complexity at most $t$.

\header{Authenticated Encryption (with Associated Data).}  In addition, to semantic security (possibly with access to an encryption/decryption oracle), we require an unforgeability property ($\ufcma$, for Unforgeability under Chosen Message Attacks). More precisely, let $\varPi=(\cE, \cD)$ be an Authenticated Encryption scheme. Consider the Experiment $\bfexp^{\ufcma}_{\varPi}(\cA)$ in which $\cA$ is given access to an encryption oracle $\cE(k,\cdot)$ and a decryption oracle $\cD(k,\cdot)$, for a random key $k\in\cK$. The Experiment returns 1 if $\cA$ outputs a valid ciphertext $c$, which means that $\cD(k,c)\neq \bot$ while $c$ has not been obtained as the output of an encryption query to $\cE(k,\cdot)$.
We define the forger's advantage of $\cA$ as $\Adv^{\ufcma}_{\varPi}(\cA)=\Pr[\bfexp^{\ufcma}_{\varPi}{\cA}=1]$ and the advantage function $\Adv^{\ufcma}_{\varPi}(t)$ as the maximum value of $\Adv^{\ufcma}_{\varPi}(\cA)$ over all $\cA$ with time-complexity at most $t$.

In the above game of basic semantic security, the adversary has no access to any encryption/decryption oracle, which is a very weak security notion, also known as one-time privacy, as the key is used once only. The One-Time Pad satisfies this property.
Note that the One-Time Pad is also injective.

\section{Key Privacy}
\label{sec:key_privacy}

An authenticated key exchange protocol \AKE can be defined using three algorithms:
\begin{itemize}
    \item $\keygen(\ID)$ takes an identity $\ID$ as input and samples a long term pair of keys ($\pk, \sk$). Key pairs are associated to that user with identity $\ID$, and the public key is added to the list $\peerpk$.
    \item $\activate(\ID, \role)$ takes as input a user identity $\ID$ and its $\role \in \{\initiator, \allowbreak \responder\}$. \activate returns a state $\state$ and a message $m'$.
    \item $\run(\ID, \sk, \peerpk, m)$ delivers $m$ to the session of user $\ID$ with secret key $\sk$ and state $\st$. $\run$ update the state $\st$ and returns the response message $m'$
\end{itemize} 
Algorithm $\run$ takes as implicit argument a state $\st$, that contains some informations, denoted in \texttt{typewriter font}, about the actual session. A value $\peerid \in \bN$ used to identify the intended partner identity of the session, a \role, either \initiator or \responder, defining the role played by the session. The state also contains the \status of the actual session. Either the session \status is \running, meaning the session has been activated, \accepted, meaning that a party has all the material to compute the session key or \terminated when the session key is computed and the protocol is ended. We also consider a \rejected flag in case the session meets a mistake in the verification phase. The final session key is stored in \SK and set as $\bot$ until defined by the key schedule. Finally, \ttsid stores the session identifier used to define partnered session in the security model.

We describe in Figure \ref{Fig:security-game} the security game introduced in \cite{EPRINT:DavGun20} following the framework by Bellare \etal \cite{EC:BelRog06}. After initializing the game, the adversary $\cA$ is given multiple access to the following queries:
\begin{itemize}
    \item \newuser: Generates a new user by generating a new pair of keys.
    \item \send: Controls activation and message processing of sessions
    \item \revSK: Reveals the session key of a terminated session.
    \item \revLTK: Corrupts a user and reveals its long term secret key.
    \item \test: Provides a real-or-random challenge on the session key of the queried session. 
\end{itemize}
Then, the adversary makes a single call to the \finalize algorithm, which returns the result of the predicate $[b'=b]$, where $b'$ is the guess of $\cA$ and $b$ is the challenge bit, after succeeding through the following predicates:

\sound: ensures at most two sessions share the same \ttsid. Once a couple of sessions is detected, the predicate checks if both of them have their \status accepted, one session user \ID is the \peerid of the other session, with different \role and the same final session key \SK. If one of these property is not verified, the adversary breaks the soundness.

\fresh: detects trivially attacked sessions. First, it ensures that neither the session key is revealed or any of the peers of the session is corrupted before the acceptance time $t_{acc}$. In a second time, the \fresh predicate ensures that the partnered session is neither tested or revealed. If such a session is detected, we set the answer bit $b'$ as 0.

The advantage of an adversary $\cA$ against the key privacy is its bias in guessing $b$, from the random choice: $\Adv^{\kp-\ake}(\cA) = \Pr[b'=b] - 1/2$.
Therefore, in Figure~\ref{Fig:EDHOC-formalized} we give a formalized description of the \EDHOC protocol compliant with the security game made in Figure~\ref{Fig:security-game}. The protocol is analyzed in the random oracle model, therefore, HKDF can be substituted by respective random oracles. 

\begin{figure}
    \begin{pchstack}[center , space=1mm] 
        \begin{pcvstack}
            
            \procedure [codesize=\scriptsize, linenumbering, jot=-1mm]{\scriptsize \initialize()}{
                \time \gets 0\\
                \users \gets 0\\
                b \sample \bin
                }
                                
                \procedure[codesize=\scriptsize, linenumbering, jot=-1mm]{\scriptsize \newuser()}{
                    \users \gets \users + 1 \\
                    (\pk_\users, \sk_\users) \sample \kgen \\
                    \revltk_\users \gets \infty \\
                    \peerpk[\users] \gets \pk_\users \\
                    \pcreturn \pk_\users        
                    }
                    
                \end{pcvstack}
            \pchspace
            \begin{pcvstack}
                
                \procedure[codesize=\scriptsize, linenumbering, jot=-1mm]{\scriptsize $\revLTK(u)$}{
                    \time \gets \time + 1 \\
                    \revltk_u \gets \time \\ 
                    \pcreturn \sk_u
                    }

                \procedure[codesize=\scriptsize, linenumbering, jot=-1mm]{\scriptsize $\revSK(u,i)$}{
                    \pcif \piui = \bot \text{ or } \pcskipln \\
                    \t \piui.\status \neq \accepted : \\
                    \t \pcreturn \bot \\
                    \piui.revealed \gets \pctrue \\
                    \pcreturn \piui.\SK 
                    }

                    \end{pcvstack}
                    \pchspace
        
            \procedure[codesize=\scriptsize, linenumbering, jot=-1mm]{\scriptsize $\finalize(b')$}{
                \pcif \neg \sound : \\
                \t \pcreturn 1 \\
                \pcif \neg \fresh : \\ 
                \t b' \gets 0 \\ 
                \pcreturn [b=b'] 
            }
    \end{pchstack}
    \begin{pcvstack}
        \begin{pchstack}
        \procedure[codesize=\scriptsize, linenumbering, jot=-1mm]{\scriptsize $\send(u,i,m)$}{
            \pcif \piui = \bot: \\
            \t (\peerid, \role) \gets m \\
            \t (\piui, m') \sample \activate(u, \sk_u, \peerid, \peerpk, \role) \\
            \t \piui . t_{acc} \gets 0 \\
            \pcelse : \\
            \t (\piui, m') \sample \run(u, \sk_u, \piui, \peerpk, \role) \\
            \pcif \piui.\status = \accepted : \\
            \t \time \gets \time + 1 \\
            \t \piui . t_{acc} \gets \time \\
            \pcreturn m'
        }

            \pchspace
                
            \procedure[codesize=\scriptsize, linenumbering, jot=-1mm]{\scriptsize $\test(u,i)$}{
                \pcif \piui = \bot \text{ or } \pcskipln \\ 
                \t \piui.\status \neq \accepted  \text{ or } \piui.\tested \\
                \t \pcreturn \bot \\
                \piui.\tested \gets \pctrue \\
                T \gets T \cup \{\piui\} \\
                k_0 \gets \piui.\SK \\
                k_1 \sample \KE.\KS\\
                \pcreturn k_b
            }
                                
        \end{pchstack}
        
        \begin{pchstack}[ space=2mm ] 
            \procedure [codesize=\scriptsize, linenumbering]{\scriptsize \sound}{
                \pcif \exists \text{ distinct } \piui, \pi_v^j, \pi_w^k \text{ with } \piui.\sid=\pi_v^j.\sid= \pi_w^k.\sid:\\
                \t \pcreturn \pcfalse \\
                \pcif \exists \piui, \pi_v^j \text{ with } \\
                \t  \piui.\status=\pi_v^j.\status=\accepted \\
                \t \text{ and } \piui.\sid=\pi_v^j.\sid \\
                \t \text{ and } \piui.\peerid=u \text{ and } \pi_v^j.\peerid=v \\
                \t \text{ and } \piui.\role \neq \pi_v^j.\role, \text{ but } \piui.\SK \neq \pi_v^j.\SK \\
                \t \pcreturn \pcfalse \\
                \pcreturn \pctrue 
            }

            \procedure[codesize=\scriptsize, linenumbering]{\scriptsize \fresh}{
                \forall \piui \in T \\
                \t \pcif \piui.\revealed  \pcskipln\\
                \t \t \text{ or } \revltk_{\piui.\peerid} < \piui.t_{acc}: \\
                \t \t \pcreturn \pcfalse \\
                \t \pcif \exists \pi_v^j \neq \piui \textbf{s.t.} \pcskipln \\
                \t \t \t \piui.\sid = \piui.\sid \text{ and } \pcskipln \\
                \t \t \t (\pi_v^j.\tested \text{ or } \pi_v^j.\revealed): \\
                \t \t \pcreturn \pcfalse \\
                \pcreturn \pctrue
            }

        \end{pchstack}

        \end{pcvstack}
    \caption{Authenticated Key Exchange Key Privacy Security Game $G_{\AKE,\cA}^{\kp-\ake}$} \label{Fig:security-game}
\end{figure}

\begin{figure}
    \begin{minipage}[t]{.5\textwidth}
        \procedure[codesize=\scriptsize, linenumbering]{\scriptsize $\keygen()$}{
            \sk \getsr \bZ_p \\
            \pk \gets g^{\bZ_p} \\
            \pcreturn (\pk, \sk)
        } \\
        \procedure [codesize=\scriptsize,linenumbering, jot=-1mm]{\scriptsize $\activate(\ID, role)$}{
            \role \gets role \\
            \status \gets \running \\
            \pcif role = \initiator : \\
            \t m' \gets \initrun1(\ID) \\
            \pcelse m' \gets \bot \\
            \pcreturn m
        }

        \procedure[codesize=\scriptsize, linenumbering, jot=-1mm]{\scriptsize $\initrun1(\ID_\sfI)$}{
            x_e \getsr \bZ_p \\
            X_e \gets g^{x_e} \\
            \sfC_\sfI \getsr \bit^{nl}  \\
            \state \gets (\sfC_\sfI, X_e, x_e) \\
            m_1 \gets  (\sfC_\sfI \| X_e \| \EAD_1) \\
            \pcreturn m_1
        } \\
        \procedure[codesize=\scriptsize, linenumbering, jot=-1mm]{\scriptsize $\initrun2(\ID_\sfI, x_s, (Y_e , c_2, \CR))$}{
            \PRK_{2e} \gets \ROT( "" , {Y_e}^{x_e}) \\
            \TH_2 \gets \cH(Y_e, \CR, \cH(m_1)) \\
             \sk_2 \gets \ROP(\PRK_{2e}, 0, \TH_2, \lPT) \\
            m_2 \gets \cD(\sk_2 , c_2) \\
            (\ID_\sfR \| t_2 \| \EAD_2) \gets m_2\\       
            \CTX_2 \gets (\ID_\sfR \| \TH_2 \| Y_s \| \EAD_2) \\
            \salt_{3e2m} \gets \ROP(\PRK_{2e}, 1, \TH_2, \lH) \\
            \PRK_{3e2m} \gets \ROT(\salt_{3e2m}, {Y_s}^{x_e}) \\
            t_2 \gets \ROP(\PRK_{3e2m} , 2, \CTX_2, \lMAC)  \\
            \pcif t'_2 \neq t_2 : \\
            \t \status \gets \rejected \\ 
            \t \pcreturn \bot \\ 
            \TH_3 \gets \cH(\TH_2 , m_2)\\
            \sk_3 \gets \ROP(\PRK_{3e2m}, 3, \TH_3, \lkey) \\ 
            \IV_3 \gets \ROP(\PRK_{3e2m}, 4,  \TH_3, \lIV) \\
            \salt_{4e3m} \gets \ROP(\PRK_{3e2m}, 5, \TH_3, \lH) \\
            \PRK_{4e3m} \gets \ROT(\salt_{4e3m}, {Y_e}^{x_s}) \\
            \status \gets \accepted \\
            \CTX_3 \gets (\ID_\sfI \| \TH_3 \| X_s \| \EAD_3) \\
            t_3 \gets \ROP(\PRK_{4e3m}, 6, \CTX_3, \lMAC) \\
            m_3 \gets (\ID_\sfI \| t_3 \| \EAD_3) \\
            c_3 \gets \cE'(\sk_3, \IV_3, m_3) \\
            \TH_4 \gets \cH(\TH_3, m_3) \\
            \PRK_\out \gets \ROP(\PRK_{4e3m},7,\TH_4, \lH) \\
            \status \gets \terminated\\
            \SK \gets \PRK_{\out} \\
            \pcreturn c_3
            }
    \end{minipage}
    \begin{minipage}[t]{.5\textwidth}
        \procedure [codesize=\scriptsize,linenumbering, jot=-1mm]{\scriptsize $\run(\ID, \sk, \peerpk, m)$}{
            \pcif \status \neq \running : \\
            \t \pcreturn \bot \\
            \pcif \role = \initiator : \\
            \t m' \gets \initrun2(\ID, \sk, m) \\
            \pcelseif \sid = \bot : \\
            \t m' \gets \resprun1(\ID, \sk, m) \\
            \pcelse : \\
            \t m' \gets \resprun2(\ID, \peerpk, m) \\
            \pcreturn m
        }
        \procedure[codesize=\scriptsize,linenumbering, jot=-1mm]{\scriptsize $\resprun1(\ID_\sfR, y_s,  m_1=(X_e, \sfC_\sfI, \EAD_1))$}{
            y_e \getsr \bZ_p \\
            Y_e \gets g^{y_e} \\
            \CR \getsr \bit^{nl}  \\
            \sid \gets (\sfC_\sfI , \CR , X_e , Y_e) \\
            \PRK_{2e} \gets \ROT( "" , {X_e}^{y_e}) \\
            \sk_2 \gets \ROP(\PRK_{2e}, 0, \TH_2, \lPT) \\
            \salt_{3e2m} \gets \ROP(\PRK_{2e}, 1, \TH_2, \lH) \\
            \PRK_{3e2m} \gets \ROT(\salt_{3e2m}, {X_e}^{y_s}) \\
            \TH_2 \gets \cH(Y_e, \CR, \cH(m_1)) \\ 
            \CTX_2 \gets (\ID_\sfR \| \TH_2 \| Y_s \| \EAD_2) \\
            t_2 \gets \ROP(\PRK_{3e2m} ,2, \CTX_2, \lMAC)  \\
            m_2 \gets (\ID_\sfR \| t_2 \| \EAD_2) \\
            c_2 \gets \cE(\sk_2, m_2) \\
            \pcreturn (Y_e , c_2, \CR)
        } \\
        \procedure[codesize=\scriptsize,linenumbering, jot=-1mm]{\scriptsize $\resprun2(\ID_\sfR, \peerpk, c_3)$}{
            \TH_3 \gets \cH(\TH_2 , m_2)\\
            \sk_3 \gets \ROP(\PRK_{3e2m}, 3, \TH_3, \lkey) \\
            \IV_3 \gets \ROP(\PRK_{3e2m}, 4, \TH_3, \lIV) \\
            m_3 \gets \cD'(\sk_3 , \IV_3, c_3) \\
            \pcif m_3 = \bot : \\
            \t \status \gets \rejected \\
            \t \pcreturn \bot \\
            (\ID_\sfI \| t_3 \| \EAD_3) \gets m_3 \\
            X_s \gets \peerpk[\ID_\sfI] \\
            \salt_{4e3m} \gets \ROP(\PRK_{3e2m}, 5, \TH_3, \lH) \\
            \PRK_{4e3m} \gets \ROT(\salt_{4e3m}, {X_s}^{y_e}) \\
            \status \gets \accepted \\
            \CTX_3 \gets (\ID_\sfI \| \TH_3 \| X_s \| \EAD_3) \\
            t'_3 \gets \ROP(\PRK_{4e3m} ,6, \CTX_3, \lMAC) \\
            \pcif t'_3 \neq t_3 : \\
            \t \status \gets \rejected \\
            \t \pcreturn \bot \\
            \TH_4 \gets \cH(\TH_3, m_3) \\
            \PRK_\out \gets \ROP(\PRK_{4e3m},7,\TH_4, \lH) \\
            \status \gets \terminated\\
            \SK \gets \PRK_{\out} \\
            \pcreturn \bot 
                }
    \end{minipage}
        \caption{Formalized description of the \EDHOC protocol} \label{Fig:EDHOC-formalized}
\end{figure}

\begin{theorem}\label{Th:KP}
    The above \EDHOC protocol satisfies the key privacy property under the Gap Diffie-Hellman problem in the Random Oracle model, and the injectivity of $(\cE,\cD)$. More precisely, with $q_\RO$ representing the global number of queries to the random oracles, $n_\sigma$ the number of running sessions, $N$ the number of users, and $\lH$ the hash digest length, we have
    $\Adv^{\kp-\ake}_{\EDHOC}(t ;  q_\RO, n_\sigma, N)$ upper-bounded by
    $$\Adv_{\bG}^{\GDH}(t, n_\sigma \cdot q_\RO)
    + 2N\cdot\Adv_{\bG}^{\GDH}(t, q_\RO)
    + \dfrac{{q_\RO}^2 +4}{2^{\lH+1}}$$
\end{theorem}

\begin{proofgame}
    \item This game is the key privacy security game $G_{\AKE,\cA}^{\kp-\ake}$ (defined in Figure~\ref{Fig:security-game}) played by $\cA$ using the \keygen, \activate and \run algorithms (defined in Figure~\ref{Fig:EDHOC-formalized}). The \keygen algorithm generates a long term pair of key, calling \activate with an user with identity $u$, $\cA$ creates its $i$-th session with $u$, denoted $\piui$. 
    $$ \Pr[\succ_0]=\Pr[G_{\AKE,\cA}^{\kp-\ake}],$$
    where the event $\succ$ means $b'=b$.

    We stress that in this security model, with Perfect Forward Secrecy, we use the weak definition of corruption, meaning that a query to \revLTK only reveals the long-term key, while the ephemeral key remains unrevealed. We say a party/session is non-corrupted if no query to \revLTK has been made before the time of acceptance $t_{acc}$, where we consider each block (\initrun1, \initrun2, \resprun1, \resprun2) as atomic. Then corruptions can only happen between two calls to simulated players. \bigskip

    \item In this game, we simulate the random oracles by lists that are empty at the beginning of the game. As $\ROT$ and $\cH$ always return a digest of size $\lH$, we simply use the simulation oracle $\SOT$ and $\SOH$ respectively. However, $\ROP$ may return values of several lengths: $\lPT$ for the one-time key encrypting the responder first message, $\lH$ for the \salt values and the session key, $\lkey$ and $\lIV$ for the AEAD key length and Initialisation Vector respectively, and $\lMAC$ for the tags. We thus define a simulation oracle by digest size: $\SOP^{\size}$, for $\size$ in $\{ \lPT, \lH, \lkey, \lIV, \lMAC \}$
    
    The simulation oracles $\SOP$ and $\SOH$ work as the usual way of simulating the answer with a new random answer for any new query, and the same answer if the same query is asked again. For the simulation oracles $\SOT$, the oracle consists in a list that contains elements of the form $(\str, Z, (X,Y) ; \lambda)$, where when first set, either $Z$ or $(X,Y)$ is non-empty. 
    Indeed, when making a call to a random oracle, the official query is of the form $(\str,Z)$, where $\str$ is any bit string, that can be empty or a pseudo-random key, and $Z$ is a Diffie-Hellman value. Then, the simulator checks in the list for an entry matching with $(\str,Z,*;\lambda)$. If such an element is found, one outputs $\lambda$, otherwise one randomly set $\lambda \getsr \bit^{\kappa}$ and append $(\str,Z,\bot;\lambda)$ to the list.
    But later, the simulator will also ask queries of the form $(\str, (X,Y))$, where $(X,Y)$ is a pair of group elements. Then one checks in the list for an entry matching with either $(\str,*,(X,Y);\lambda)$ or $(\str,Z,*;\lambda)$ such that $\DDH(g,X,Y,Z)=1$. If such an element is found, one outputs $\lambda$, otherwise one randomly set $\lambda \getsr \bit^{\kappa}$ and append $(\str,\bot,(X,Y);\lambda)$ to the list. 
    When such new kinds of elements exist in the list, for the first kind of queries $(\str, Z)$, one checks in the list for an entry matching with either $(\str,Z,*;\lambda)$ as before, or $(\str,*,(X,Y);\lambda)$ such that $\DDH(g,X,Y,Z)=1$. 
    We detail in Figure \ref{game:lists} the functioning of those oracles, and the modifications made to the simulation.

\begin{figure}[htb]
    \begin{pchstack}
       \begin{pcvstack}
           
        \procedure[codesize=\scriptsize, linenumbering, jot=-1mm]{\scriptsize $\SOT(\str, \sfinput)$}{
            \pcif \len(\sfinput)=1 : \\
            \t Z \gets \sfinput \\
            \t \pcif \exists (\str,Z, * ; \lambda) \in \SOT  : \\
            \t \t  \pcreturn \lambda \\
            \t \pcelse : \\
            \t \t \pcif \exists (\str, \bot, (X,Y); \lambda) \in \SOT \pcskipln\\
            \t \t \t \t \t \text{ s.t. } \DDH(X,Y,Z)=1 : \pcskipln \\
            \t \t  \pccomment{update $\SOT$} \\
            \t \t \t \SOT^{-1}[\lambda] \gets (\str, Z, (X,Y); \lambda)\\
            \t \t \t \pcreturn \lambda \\
            \t \t \pcelse : \\
            \t \t \t \lambda \getsr \bit^{\kappa} \\
            \t \t \t \SOT \gets \SOT \cup \{(\str , Z , \bot ; \lambda)\} \\
            \t \t \t \pcreturn \lambda \\
            \pcelse: \pcskipln \\
            \pccomment{$\sfinput=(X,Y)$, only by the simulator} \\
            \t (X,Y) \gets \sfinput  \\
            \t \pcif \exists (\str, *, (X,Y) ; \lambda) \in \SOT  : \\
            \t \t  \pcreturn \lambda \\
            \t \pcelse : \\
            \t \t \pcif \exists (\str, Z, \bot; \lambda) \in \SOT \pcskipln \\
            \t \t \t \t \t \text{ s.t. } \DDH(X,Y,Z)=1: \\
            \t \t \t \SOT^{-1}[\lambda] \gets (\str, Z, (X,Y); \lambda)\\
            \t \t \t \pcreturn \lambda \\
            \t \t \pcelse : \\
            \t \t \t \lambda \getsr \bit^{\kappa}  \\
            \t \t \t \SOT \gets \SOT \cup \{(\str , \bot , (X,Y) ; \lambda)\} \\
            \t \t \t \pcreturn \lambda 
            }
            
            \procedure[codesize=\scriptsize, lnstart=4,linenumbering, jot=-1mm]{\scriptsize $\resprun1(\ID_\sfR, y_s, m)$}{
                \PRK_{2e} \gets \SOT( "" , {X_e}^{y_e}) \\
                \sk_2 \gets \SOP(\PRK_{2e}, 0, \TH_2, \lPT) \\
                \salt_{3e2m} \gets \SOP(\PRK_{2e}, 1, \TH_2, \lH) \\
                \PRK_{3e2m} \gets \SOT(\salt_{3e2m}, {X_e}^{y_s}) \\
                \TH_2 \gets \SOH(Y_e, \CR, \SOH(m_1))   \pcskiptoln{11} \\ 
                t_2 \gets \SOP(\PRK_{3e2m} ,2, \CTX_2, \lMAC)  
                }
        \end{pcvstack} 
            \pchspace 
            \begin{pcvstack}
                
                \procedure[codesize=\scriptsize,linenumbering, jot=-1mm]{\scriptsize $\initrun2(\ID_\sfI, x_s, m)$}{
                    \PRK_{2e} \gets \SOT( "" , {Y_e}^{x_e}) \\
                    \TH_2 \gets \SOH(Y_e, \CR, \SOH(m_1)) \\
                    \sk_2 \gets \SOP(\PRK_{2e}, 0, \TH_2, \lPT)  \pcskiptoln{7} \\
                    \salt_{3e2m} \gets \SOP(\PRK_{2e}, 1, \TH_2, \lH) \\
                    \PRK_{3e2m} \gets \SOT(\salt_{3e2m}, {Y_s}^{x_e}) \\
                    t_2 \gets \SOP(\PRK_{3e2m} , 2, \CTX_2, \lMAC)   \pcskiptoln{13} \\
                    \TH_3 \gets \SOH(\TH_2 , m_2)\\
                    \sk_3 \gets \SOP(\PRK_{3e2m}, 3, \TH_3, \lkey) \\ 
                    \IV_3 \gets \SOP(\PRK_{3e2m}, 4,  \TH_3, \lIV) \\
                    \salt_{4e3m} \gets \SOP(\PRK_{3e2m}, 5, \TH_3, \lH) \\
                    \PRK_{4e3m} \gets \SOT(\salt_{4e3m}, {Y_e}^{x_s})  \pcskiptoln{20} \\
                    t_3 \gets \SOP(\PRK_{4e3m}, 6, \CTX_3, \lMAC)  \pcskiptoln{23} \\
                    \TH_4 \gets \SOH(\TH_3, m_3) \\
                    \PRK_\out \gets \SOP(\PRK_{4e3m},7,\TH_4, \lH)
                    }
            
                    \procedure[codesize=\scriptsize,linenumbering, jot=-1mm]{\scriptsize $\resprun2(\ID_\sfR, \peerpk, c_3)$}{
                        \TH_3 \gets \SOH(\TH_2 , m_2)\\
                        \sk_3 \gets \SOP(\PRK_{3e2m}, 3, \TH_3, \lkey) \\
                        \IV_3 \gets \SOP(\PRK_{3e2m}, 4, \TH_3, \lIV) \pcskiptoln{10} \\
                        \salt_{4e3m} \gets \SOP(\PRK_{3e2m}, 5, \TH_3, \lH) \\
                        \PRK_{4e3m} \gets \SOT(\salt_{4e3m}, \peerpk[\ID_\sfI]^{y_e}) \pcskiptoln{14} \\
                        t'_3 \gets \SOP(\PRK_{4e3m} ,6, \CTX_3, \lMAC) \pcskiptoln{18} \\
                        \TH_4 \gets \SOH(\TH_3, m_3) \\
                        \PRK_\out \gets \SOP(\PRK_{4e3m},7,\TH_4, \lH) 
                    }
                \end{pcvstack}
                            
                        \end{pchstack}  
                        \caption{Description of $\SOT$ list queries and modifications to the simulation} \label{game:lists}
                    \end{figure}
                    
        Thanks to the \DDH oracle, this simulation is perfect, and is thus indistinguishable to the adversary:
        $$ \Pr[\succ_{\thisgame}]=\Pr[\succ_{\thepreviousgame}] $$
                    
    \item In order to prevent collisions in the future \PRK generation, we modify the simulation oracles $\SOT, \SOP^\lH$ and $\SOH$, such that if a collision occurs, the simulator stops. We therefore need to determine the probability of a collision, to bound the probability for an adversary to distinguish this game from the previous one. To do so, we rely on the birthday paradox. By denoting $q_{\SOT}, q_{\SOP^\lH}, q_{\SOH}$ the amount of queries made to oracles $\SOT, \SOP^\lH, \SOH$ respectively, the birthday paradox bound gives: 
    $$ \Pr[\succ_{\thisgame}] - \Pr[\succ_{\thepreviousgame}] \leq \dfrac{{q_\SOT}^2 + {q_{\SOP^\lH}}^2 + {q_\SOH}^2}{2^{\lH+1}} $$
    
    \item One can note that thanks to the above simulation of the random oracles, the simulator does not need anymore to compute Diffie-Hellman values. Then, for every simulated player, the simulator generates $X_e$ or $Y_e$ at random in the group, and the simulation is still performed as in the previous game. As corruption queries only reveal long-term secret, still known to the simulator, the view of the adversary is perfectly indistinguishable of the previous game and we have: 
    $$ \Pr[\succ_{\thisgame}]=\Pr[\succ_{\thepreviousgame}] $$

    \item In this game, when simulating any \textbf{initiator} receiving a forged tuple $(Y_e,c_2,\CR)$ from the adversary in the name of a \textbf{non-corrupted user}, one simulates $\PRK_{3e2m}$ thanks to a private oracle $\SOK{3e2m}$, which makes it perfectly unpredictable to the adversary.
    If the pair $(Y_e, \CR)$ is forged, $\TH_2$ and $\salt_{3e2m}$ are different from the values obtained by a possibly simulated responder, thanks to the absence of collisions as they are respectively computed using $\SOH$ and $\SOP^\lH$.
    Otherwise, $\sk_2$ is not modified. So if the ciphertext $c_2$ is forged, thanks to the injective property of the encryption scheme $(\cE,\cD)$ when the key is fixed, $m_2$, and by consequent $\TH_3$ and $\salt_{4e3m}$ are different from the values obtained by a possibly simulated responder. 
    In order to detect the inconsistency of $\PRK_{3e2m}$ with respect to the public oracle answer, the adversary must have  asked $\SOT$ on the correct Diffie-Hellman value ${X_e}^{y_s}$.
    We denote the event $F_1$, that query ${X_e}^{y_s}$ is asked whereas $y_s$ is the long-term secret key of a non-corrupted user and $X_e$ has been generated by the simulator. If this event happens (which can easily be checked as the simulator knows $y_s$), one stops the simulation:
    $$|\Pr[\succ_{\thisgame}]-\Pr[\succ_{\thepreviousgame}]| \leq \Pr[F_1].$$

    \item[\thegames'] We now provide an upper-bound on $\Pr[F_1]$: given a \GDH challenge $(X=g^x,Y=g^y)$, one simulates all the $X_e$ as $X_e=X\cdot g^r$, for random $r \getsr \bZ_p$, but chooses one user to set $Y_s=Y$. Even if $y_s$ is therefore not known, simulation is still feasible as the simulator can make query to the $\SOT$ oracle with input $(X_e, Y_s)$.  Then, one can still answer all the corruption queries, excepted for that user. But anyway, if $F_1$ happens on that user, this user must be non-corrupted at that time: one has solved the \GDH problem, and one can stop the simulation. If the guess on the user is incorrect, one can also stop the simulation:
    $\Pr[F_1] \leq N\cdot\Adv_{\bG}^{\GDH}(t, q_\RO)$, where $N$ is the number of users in the system.

    \item In this game, when simulating any \textbf{responder} receiving a forged message $m_1$ from the adversary in the name of a \textbf{non-corrupted user}, still non-corrupted when sending $c_3$ to $\resprun2$, one simulates $\PRK_{4e3m}$ thanks to a private oracle $\SOK{4e3m}$, which makes it perfectly unpredictable to the adversary.
    Since $m_1$ is forged, thanks to the absence of collisions, $\TH_2$,$\TH_3$, and $\salt_{4e3m}$ are different from the values obtained by a possibly simulated responder.
    In order to detect the inconsistency of $\PRK_{4e3m}$ with respect to the public oracle answer, the adversary must have  asked $\SOT$ on the correct Diffie-Hellman value ${Y_e}^{x_s}$.
    We denote the event $F_2$, that query ${Y_e}^{x_s}$ is asked whereas $x_s$ is the long-term secret key of a non-corrupted user and $Y_e$ has been generated by the simulator. If this event happens (which can easily be checked as the simulator knows $x_s$), one stops the simulation:
    $$|\Pr[\succ_{\thisgame}]-\Pr[\succ_{\thepreviousgame}]| \leq \Pr[F_2].$$
    
    \item[\thegames'] We now provide an upper-bound on $\Pr[F_2]$: given a \GDH challenge $(X=g^x,Y=g^y)$, one simulates all the $Y_e$ as $Y_e= Y \cdot g^{r'}$, for random ${r'} \getsr \bZ_p$, but chooses one user to set $X_s=X$. Then, one can still answer all the corruption queries, excepted for that user. But anyway, if $F_2$ happens on that user, this user must be non-corrupted at that time: one has solved the \GDH problem, and one can stop the simulation. If the guess on the user is incorrect, one can also stop the simulation:
    $\Pr[F_2] \leq N\cdot\Adv_{\bG}^{\GDH}(t, q_\RO)$.

    \item In this game, we simulate the key generation of $\PRK_{2e}$, for all the passive sessions ($m_1$ received by a simulated responder comes from a simulated initiator, or $(Y_e, c_2, \CR)$ received by a simulated initiator comes from a simulated responder, and both used the same $m_1$ as first message), thanks to a private oracle $\SOK{2e}$, acting in the same vein as $\SOT$, but not available to the adversary.
    This makes a difference with the previous game if the key $\PRK_{2e}$ has also been generated by asking $\SOT$ on the correct Diffie-Hellman value $Z=g^{x_e y_e}$. We denote by $F_3$ the latter event, and stop the simulation in such a case:
    $$|\Pr[\succ_{\thisgame}]-\Pr[\succ_{\thepreviousgame}]| \leq \Pr[F_3]$$

    \item[\thegames'] We now provide an upper-bound on $\Pr[F_3]$. Given a \GDH challenge $(X=g^x,Y=g^y)$, one simulates all the $X_e$ as $X_e= X \cdot g^{r}$, for random ${r} \getsr \bZ_p$, and all the $Y_e$ as $Y_e= Y \cdot g^{r'}$, for random ${r'} \getsr \bZ_p$. As the key $\PRK_{2e}$ does not depend on the session context, any query $Z$ to the $\SOT$ oracle can make $F_3$ occurs on any of the passive session pairs $(X_e = X \cdot g^{r},Y_e = Y \cdot g^{r'})$, we upper-bound the number by $n_\sigma$. Hence, $q_\RO$ \DDH-oracle queries might be useful to detect $F_3$ on an input $Z = \CDH(X_e,Y_e) = g^{xy} \cdot X^{r'} \cdot Y^{r} \cdot g^{rr'}$, solving the \GDH challenge $(X,Y)$:
    $$\Pr[F_3] \leq \Adv_{\bG}^{\GDH}(t, n_\sigma \cdot q_\RO).$$

    \item In this game, when simulating any \textbf{initiator} receiving the second message $(Y_e , c_2, \CR)$, from the adversary in the name of a \textbf{non-corrupted user}, one simulates $\PRK_{3e2m}$ thanks to a private oracle $\SOK{3e2m}$.
    This makes a difference with the previous game only if this is a passive session, in which case $\PRK_{2e}$ is unpredictable, and thus different from the public one excepted with probability $2^{-\lH}$.
    As there are no collision, $\salt_{3e2m}$ is different from the value obtained by a possibly simulated responder.
    In order to detect the inconsistency of $\PRK_{3e2m}$ with respect to the public oracle answer, the adversary must have  asked $\SOT$ on the correct Diffie-Hellman value $X_e^{y_s}$, which is not possible as event $F_1$ stops the simulation:
    $$|\Pr[\succ_{\thisgame}]-\Pr[\succ_{\thepreviousgame}]| \leq \dfrac{1}{2^{\lH}}.$$
    
    \item In this game, when simulating any \textbf{initiator} receiving the second message $(Y_e , c_2, \CR)$, from the adversary in the name of a \textbf{non-corrupted user}, one simulates $\PRK_{4e3m}$ thanks to a private oracle $\SOK{4e3m}$.
    In this case, $\PRK_{3e2m}$ is unpredictable, as well as $\salt_{4e3m}$ and $\PRK_{4e3m}$:
    $$\Pr[\succ_{\thisgame}] = \Pr[\succ_{\thepreviousgame}].$$

    \item In this game, when simulating any \textbf{responder} receiving $c_3$, from the adversary in the name of a \textbf{non-corrupted user}, one simulates $\PRK_{4e3m}$ thanks to the private oracle $\SOK{4e3m}$.
    This makes a difference with the previous game only if this is not a passive session, in which case $\PRK_{2e}$ is unpredictable, and thus different from the public one excepted with probability $2^{-\lH}$.
    As there are no collision, $\salt_{3e2m}$, $\PRK_{3e2m}$, and $\salt_{4e3m}$ are different from the values obtained by a possibly simulated responder.
    In order to detect the inconsistency of $\PRK_{4e3m}$ with respect to the public oracle answer, the adversary must have asked $\SOT$ on the correct Diffie-Hellman value $Y_e^{x_s}$, which is not possible as event $F_2$ stops the simulation:
    $$|\Pr[\succ_{\thisgame}]-\Pr[\succ_{\thepreviousgame}]| \leq \dfrac{1}{2^{\lH}}.$$

    \item In this game, for any fresh session, one simulates $\PRK_{\out}$ thanks to the private oracle $\SOK{\out}$.
    A session being fresh means that no corruption of the party or of the partner occurred before the time of acceptance: the initiator is not corrupted before receiving $(Y_e,c_2,\CR)$ and the responder is not corrupted before receiving $c_3$. By consequent, they are not corrupted before $\PRK_{4e3m}$ was computed.
    We have seen above that in those cases, the key $\PRK_{4e3m}$ is generated using the private oracle $\SOK{4e3m}$: it is unpredictable. The use of the private oracle $\SOK{\out}$ can only be detected if the query $\PRK_{4e3m}$ is asked to $\SOP$:
    $$|\Pr[\succ_{\thisgame}]-\Pr[\succ_{\thepreviousgame}]| \leq \dfrac{q_{\SOP^\lH}}{2^{\lH}}.$$
    
    Globally, one can note that the gap between the initial and the last games is upper-bounded by
    \begin{align*}
        \Adv_{\bG}^{\GDH}(t, n_\sigma \cdot q_\RO)
        & + 2N\cdot\Adv_{\bG}^{\GDH}(t, q_\RO) \\
        & + \dfrac{{q_\SOT}^2 + {q_{\SOP^\lH}}^2 + {q_\SOH}^2}{2^{\lH+1}} + \dfrac{2 + q_{\SOP^\lH}}{2^{\lH}} \\
        \leq  \Adv_{\bG}^{\GDH}(t, n_\sigma \cdot q_\RO)
        & + 2N\cdot\Adv_{\bG}^{\GDH}(t, q_\RO)
        + \dfrac{{q_\RO}^2 + 4}{2^{\lH+1}}
    \end{align*}
    Eventually, for all the fresh sessions, in the real case $(b=0)$, the private oracle is used, and  outputs a random key, while in the random case $(b=1)$, the session key is random too:
    $$\Pr[\succ_\thisgame]=\frac{1}{2}.$$
    This concludes the proof.

                    \end{proofgame}

\section{Explicit Authentication}
\label{sec:EA}

Explicit authentication (or mutual authentication) aims to ensure each participant has the material to compute the final session key (accepts) when the partner terminates.
In the \EDHOC protocol, this means the responder (resp. the initiator) owns the private long-term key $y_s$ (resp $x_s$) associated to the long-term public key $Y_s$ (resp. $X_s$), and the private ephemeral keys, when the partner terminates.
\begin{figure}[htb]
    
    \procedure[codesize=\scriptsize, linenumbering]{\scriptsize \finalize}{
        \pcreturn: \pcskipln\\  
        \forall \piui \text{ s.t. }
        \begin{cases}
            \piui.\status=\terminated \tabularnewline 
            \piui.t_{acc} < \revltk_{\piui.\peerid}
        \end{cases} , 
        \exists \pi_v^j \ \text{ s.t. }\begin{cases} \piui.\peerid = v \tabularnewline
            \pi_v^j.\peerid=u \tabularnewline
            \piui.\sid = \pi_v^j.\sid \tabularnewline
            \piui.\role \neq \pi_v^j.\role \tabularnewline 
            \pi_v^j.\status = \accepted
        \end{cases}
        }

    \caption{Finalize Function for the Explicit Authentication Security Game} \label{Fig:ExpAuth-Finalize} 
\end{figure}

To do so, the responder uses $y_s$ in \resprun1 to compute $\PRK_{3e2m}$ used for the tag $t_2$ and the key $\sk_3$.
In the same way, the initiator uses $x_s$ to compute $\PRK_{4e3m}$, used for the tag $t_3$. Furthermore, they both have to use their ephemeral keys to compute $\PRK_{2e}$, used for $\sk_2$.

\paragraph{Responder Authentication.} Consider a simulated \textbf{initiator} receiving a forged message $(Y_e,c_2,\CR)$ from the adversary in the name of a \textbf{non-corrupted user}. In such a case, consider the modifications made in the key privacy proof up to the game \game7. Hence, we have replaced the generation of $\PRK_{3e2m}$ with a private oracle. Then the advantage of the adversary in breaking the explicit authentication of the responder in this game is bounded by $\dfrac{1}{2^{\lMAC}}$, added to the gap induced by the modifications made up to the game \game7. This leads to the following theorem:

\begin{theorem}\label{Th:Resp-Auth}
    The above \EDHOC protocol satisfies the responder authentication under the Gap Diffie-Hellman problem in the Random Oracle model, and the injectivity of $(\cE,\cD)$. More precisely, with $q_\RO$ representing the global number of queries to the random oracles, $n_\sigma$ the number of running sessions, $N$ the number of users, and $\lH$ the hash digest length, we have
    $\Adv^{\auth-\resp}_{\EDHOC}(t ;  q_\RO, n_\sigma, N)$ upper-bounded by
    $$\Adv_{\bG}^{\GDH}(t, n_\sigma \cdot q_\RO)
    + 2N\cdot\Adv_{\bG}^{\GDH}(t, q_\RO)
    + \dfrac{{q_\RO}^2 +2}{2^{\lH+1}} + \frac{1}{2^{\lMAC}}$$
\end{theorem}
\header{Optimal Reduction.}
One cannot expect more after these three flows, as the adversary can play the role of the responder with known $y_e$. Without knowing $y_s$, it just gets stuck to compute $\PRK_{3e2m}$ and thus $t_2$. But it can guess it (with probability $2^{-\lMAC}$), breaking authentication. But it will not know $\SK$.

\paragraph{Initiator Authentication.} We now consider any \textbf{responder} receiving a forged message $c_3$ from the adversary in the name of a \textbf{non-corrupted user}.
As above, considering the modifications made in the key privacy proof up to the game \game8, we have replaced the generation of $\PRK_{4e3m}$ with a private oracle. Then the advantage of the adversary in breaking the explicit authentication of the initiator in this game is bounded by $\dfrac{1}{2^{\lMAC}}$, added to the gap induced by the modifications made up to the game \game7. This leads to the following theorem:
\begin{theorem}\label{Th:Init-Auth}
    The above \EDHOC protocol satisfies the initiator authentication under the Gap Diffie-Hellman problem in the Random Oracle model, and the injectivity of $(\cE,\cD)$. More precisely, with $q_\RO$ representing the global number of queries to the random oracles, $n_\sigma$ the number of running sessions, $N$ the number of users, and $\lH$ the hash digest length, we have
    $\Adv^{\auth-\init}_{\EDHOC}(t ;  q_\RO, n_\sigma, N)$ upper-bounded by
    $$\Adv_{\bG}^{\GDH}(t, n_\sigma \cdot q_\RO)
    + 2N\cdot\Adv_{\bG}^{\GDH}(t, q_\RO)
    + \dfrac{{q_\RO}^2 +4}{2^{\lH+1}} + \frac{1}{2^{\lMAC}}$$
\end{theorem}
\header{Optimal Reduction.}
One cannot expect more after these three flows, as the adversary can play the role of the initiator with known $x_e$. Without knowing $x_s$, it just gets stuck to compute $\PRK_{4e3m}$ and thus $t_3$. But it can guess it (with probability $2^{-\lMAC}$) and encrypt it, as it knows $\sk_3$, breaking authentication. But it will not know $\SK$.

\section{Identity Protection}
\label{sec:IP}

Let us now consider anonymity, with identity protection.
More precisely, we want to show that the initiator's identity ($\ID_\sfI$) is protected against active adversaries, while responder's identity ($\ID_\sfR$) is protected only against passive adversaries. 

The values $\ID_\sfI$ and $\ID_\sfR$ are the authentication credentials containing the public authentication keys of I and R, respectively.

\paragraph{Responder's Identity Protection.} 
The value $\ID_\sfR$ is used in the computation of $\CTX_2$ itself used to compute $t_2$, which together with $\ID_\sfR$ constitute the first part of $m_2=(\ID_\sfR \| t_2 \| \EAD_2)$ whose encryption is $c_2$ under $\sk_2$. 
For the sake of clarity, we set $\EAD_2=$"" as $\EAD_2$ is independent from the identity of the responder and has no cryptographic purpose.  
As a responder, the passive adversary can only earn information about $\ID_\sfR$ using the ciphertext $c_2$. 
We thus define the responder identity protection experiment as follows: 
\begin{pchstack}[center]
        \procedure[linenumbering, jot=-1mm]{$\exp^{\ID-\resp-b}_{\EDHOC}$}{
        \ID_{\sfR_0} , \ID_{\sfR_1} \gets \cA(\peerid) \\
        m_1 \gets \cA(\initrun1(.)) \\
        b \gets \bin \\
        \ID_\sfR \gets \ID_{\sfR_b} \\
        y_s \gets \sk_{\ID_\sfR} \\
        (Y_e, c_2, \CR) \gets \resprun1(\ID_\sfR ,y_s, m_1) \\
        b' \gets \cA(c_2) \\
        \pcreturn b = b' }
\end{pchstack}
We define the advantage $\Adv^{\ID-\resp-b}_{\EDHOC}$ of the adversary in breaking the responder mutual authentication of \EDHOC by:
$$ \Adv^{\ID-\resp-b}_{\EDHOC}(t) = | \Pr[\exp^{\ID-\resp-0}_{\EDHOC}=1] - \Pr[\exp^{\ID-\resp-1}_{\EDHOC}=1]|$$
\begin{theorem}\label{Th:Resp-IP}
    The above \EDHOC protocol protects Responder's Identity under the Gap Diffie-Hellman problem in the Random Oracle model, the injectivity and the semantic security of $\varPi=(\cE,\cD)$. More precisely, with $q_\RO$ representing the global number of queries to the random oracles, $n_\sigma$ the number of running sessions, $N$ the number of users, and $\lH$ the hash digest length, we have
    $\Adv^{\ID-\resp-b}_{\EDHOC}(t ;  q_\RO, n_\sigma, N)$ upper-bounded by
    $$\Adv_{\bG}^{\GDH}(t, n_\sigma \cdot q_\RO)
    + 2N\cdot\Adv_{\bG}^{\GDH}(t, q_\RO)
    + \Adv_{\varPi}^{\ind}(t)
    + \dfrac{{q_\RO}^2+2}{2^{\lH+1}}$$
\end{theorem}

\begin{proofgame}
    \item This game is $\exp^{\ID-\resp-0}_{\EDHOC}$.
    The simulated initiator follows the protocol, computes $c_2=\cE(\sk_2,(\ID_\sfR \| t_2 ))$ and sends $Y_e, c_2, C_R$ to the adversary:
    $$\Pr[\succ_\thisgame]=\Pr[\exp^{\ID-\resp-0}_{\EDHOC}=1]$$
    
    \item In this game, we applied the modification made from \game0 up to \game6 in the key privacy proof. 
   \begin{align*}
        |\Pr[\succ_\thisgame]-\Pr[\succ_\thepreviousgame]| 
        & \leq \Adv_{\bG}^{\GDH}(t, n_\sigma \cdot q_\RO) \\
        & + 2N\cdot\Adv_{\bG}^{\GDH}(t, q_\RO)
        + \dfrac{{q_\RO}^2}{2^{\lH+1}}
    \end{align*}
    \item In this game, one simulates $\sk_2$ thanks to a private oracle, which makes a difference only if the random $\PRK_{2e}$ is asked to the public oracle:
    $$|\Pr[\succ_\thisgame]-\Pr[\succ_\thepreviousgame]| \leq \dfrac{1}{2^{\lH}}$$
    \item In this game, we replace the line 4 of the experiment by $\ID_\sfR \gets \ID_{\sfR_{1-b}}$, leading to the instanciation of $\exp^{\ID-\resp-1}_{\EDHOC}$.
    As $\sk_2$ is chosen at random, using the semantic security of the encryption scheme $(\cE,\cD)$, we thus have
    $$|\Pr[\succ_\thisgame]-\Pr[\succ_\thepreviousgame]| \leq \Adv_{\varPi}^{\ind}(t) $$
\end{proofgame}

\paragraph{Initiator's Identity Protection.} In this case, we expect an active security: we consider the simulation of an initiator interacting with an adversary playing in the name of a non-corrupted user with public long-term key $Y_s$. We have simulated $\PRK_{3e2m}$ with a private oracle, which leads to a private random key $\sk_3$, unless the query has been asked, with the same argument as above.

The value $\ID_\sfI$ is used in the computation of $\CTX_3$ itself used to compute $t_3$, which together with $\ID_\sfI$ constitute the first part of the message $m_3=(\ID_\sfR \| t_2 \| \EAD_2)$ whose encryption is $c_3$ under $\sk_3$. 
As above, for the sake of generality, we set $\EAD_3=$"". 
One can note that the first message $m_1$ sent by the initiator is independent of $\ID_\sfR$. We therefore start the experiment after the adversary sent his first message $(Y_e, c_2, \CR)$:
\begin{pchstack}[center]
    \procedure[linenumbering, jot=-1mm]{$\exp^{\ID-\init-b}_{\EDHOC}$}{
        \ID_{\sfI_0} , \ID_{\sfI_1} \gets \cA(\peerid) \\
        (Y_e, c_2, \CR) \gets \cA(\resprun1(.)) \\
        b \gets \bin \\
        \ID_\sfI \gets \ID_{\sfI_b} \\
        x_s \gets \sk_{\ID_\sfI} \\
        Y_s \gets  \peerpk[\ID_\sfI] \\
        c_3 \gets \initrun2(\ID_\sfI ,x_s, Y_s,(Y_e, c_2, \CR)) \\
        b' \gets \adv(c_3) \\
        \pcreturn b = b' }
\end{pchstack}
We define the advantage $\Adv^{\ID-\init-b}_{\EDHOC}$ of the adversary in breaking the responder mutual authentication of \EDHOC by:
$$ \Adv^{\ID-\init-b}_{\EDHOC}(t) = | \Pr[\exp^{\ID-\init-0}_{\EDHOC}=1] - \Pr[\exp^{\ID-\init-1}_{\EDHOC}=1]|$$

\begin{theorem}\label{Th:Init-IP}
    The above \EDHOC protocol protects Initiator's Identity under the Gap Diffie-Hellman problem in the Random Oracle model, the injectivity of $(\cE,\cD)$ and the semantic security of $\varPi'=(\cE',\cD')$. More precisely, with $q_\RO$ representing the global number of queries to the random oracles, $n_\sigma$ the number of running sessions, $N$ the number of users, and $\lH$ the hash digest length, we have
    $\Adv^{\ID-\init-b}_{\EDHOC}(t ;  q_\RO, n_\sigma, N)$ upper-bounded by
    $$\Adv_{\bG}^{\GDH}(t, n_\sigma \cdot q_\RO)
    + 2N\cdot\Adv_{\bG}^{\GDH}(t, q_\RO)
    + \Adv_{\varPi'}^{\ind}(t)
    + \dfrac{{q_\RO}^2+2}{2^{\lH+1}}$$
\end{theorem}

\section{Improvements}
\label{sec:improvements}

\begin{figure}[p]
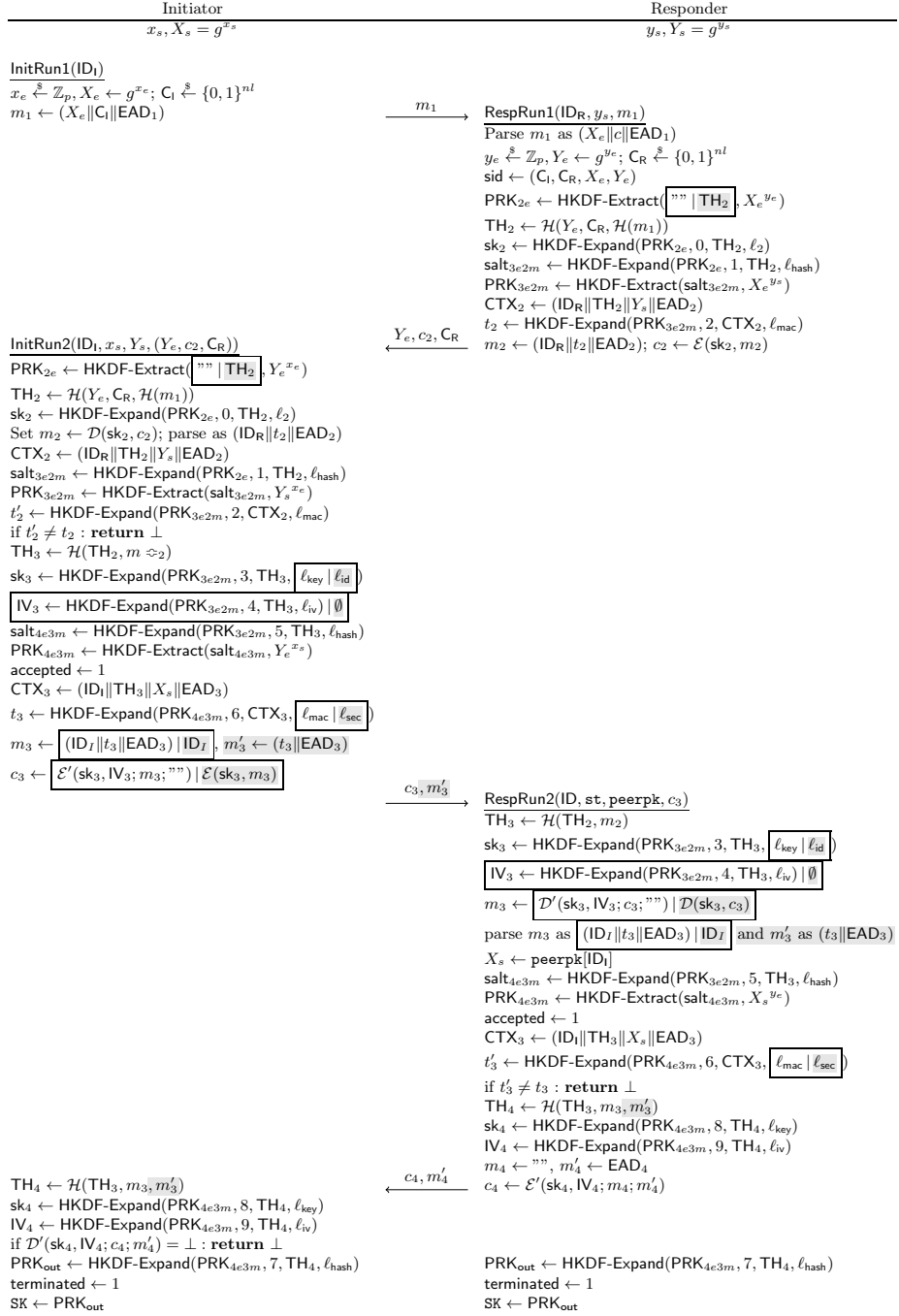

    \resizebox{\textwidth}{!}{
    \small
    \begin{tabular}{lcl}
       \multicolumn{1}{c}{Initiator} & & \multicolumn{1}{c}{Responder} \\ \hline 
       \multicolumn{1}{c}{$x_s, X_s = g^{x_s}$} & & \multicolumn{1}{c}{$y_s, Y_s = g^{y_s}$} \\
        && \\
        $\underline{\initrun1(\ID_\sfI )}$ && \\
        $x_e \getsr \bZ_p, X_e \gets g^{x_e}$; 
        $\CI \getsr \bit^{nl} $ \\
        $m_1 \gets (X_e \| \CI \| \EAD_1)$ 

        & \sendright{$m_1$}{16mm}  & $\underline{\resprun1(\ID_\sfR, y_s, m_1)}$ \\

        &&  Parse $m_1$ as $(X_e \| c \| \EAD_1)$  \\
        && $y_e \getsr \bZ_p, Y_e \gets g^{y_e}$;
        $\CR \getsr \bit^{nl} $ \\
        && $ \sid \gets (\CI , \CR , X_e , Y_e)$ \\
        && $\PRK_{2e} \gets \HKDFExt( \oldnew{""}{\TH_2} , {X_e}^{y_e})$ \\
        && $\TH_2 \gets \cH(Y_e, \CR, \cH(m_1))$ \\ 
        && $\sk_2 \gets \HKDFExp(\PRK_{2e}, 0, \TH_2, \lPT)$ \\
        && $\salt_{3e2m} \gets \HKDFExp(\PRK_{2e}, 1, \TH_2, \lH)$ \\
        && $\PRK_{3e2m} \gets \HKDFExt(\salt_{3e2m}, {X_e}^{y_s})$ \\
        && $\CTX_2 \gets (\ID_\sfR \| \TH_2 \| Y_s \| \EAD_2)$ \\
        && $t_2 \gets \HKDFExp(\PRK_{3e2m} , 2, \CTX_2, \lMAC) $ \\
        
        $\underline{\initrun2(\ID_\sfI, x_s, Y_s,(Y_e, c_2, \CR) )}$ & \sendleft{$Y_e , c_2, \CR$}{16mm} & $m_2 \gets (\ID_\sfR \| t_2 \| \EAD_2 )$; $c_2 \gets\cE(\sk_2 , m_2)$ \\
        

        $\PRK_{2e} \gets \HKDFExt( \oldnew{""}{\TH_2} , {Y_e}^{x_e})$ \\
        $\TH_2 \gets \cH(Y_e, \CR, \cH(m_1))$ \\
        $ \sk_2 \gets \HKDFExp(\PRK_{2e}, 0, \TH_2, \lPT)$ \\
        Set $m_2 \gets \cD(\sk_2 , c_2)$; parse as $(\ID_\sfR \| t_2 \| \EAD_2)$ \\       
        $\CTX_2 \gets (\ID_\sfR \| \TH_2 \| Y_s \| \EAD_2)$ \\
        $\salt_{3e2m} \gets \HKDFExp(\PRK_{2e}, 1, \TH_2, \lH)$ \\
        $\PRK_{3e2m} \gets \HKDFExt(\salt_{3e2m}, {Y_s}^{x_e})$ \\
        $t'_2 \gets \HKDFExp(\PRK_{3e2m} , 2, \CTX_2, \lMAC) $ \\
        if $t'_2 \neq t_2$ : \textbf{return} $\bot$ \\ 
        $\TH_3 \gets \cH(\TH_2 , m\Bumpeq _2)$\\
        $\sk_3 \gets \HKDFExp(\PRK_{3e2m}, 3, \TH_3, \oldnew{\lkey}{\lid} )$ \\ 
        $\oldnew{\IV_3 \gets \HKDFExp({\PRK_{3e2m}}, 4,  \TH_3, \lIV)}{\emptyset}$ \\
        $\salt_{4e3m} \gets \HKDFExp(\PRK_{3e2m}, 5, \TH_3, \lH)$ \\
        $\PRK_{4e3m} \gets \HKDFExt(\salt_{4e3m}, {Y_e}^{x_s})$ \\
        $\accepted \gets 1$ \\
        $\CTX_3 \gets (\ID_\sfI \| \TH_3 \| X_s \| \EAD_3)$ \\
        $t_3 \gets \HKDFExp(\PRK_{4e3m}, 6, \CTX_3, \oldnew{\lMAC}{\lSEC})$ \\
        $m_{3} \gets \oldnew{(\ID_I \| t_3 \| \EAD_3)}{\ID_I}$, \hl{$m'_3 \gets ( t_3 \| \EAD_3 )$} \\
        $c_{3} \gets \oldnew{\cE'(\sk_3, \IV_3; m_3;"")}{\cE(\sk_3, m_3)}$ \\
        & \sendright{$c_3 \mathhl{, m'_3}$}{16mm} &  $\underline{\resprun2(\ID, \st, \peerpk, c_3)}$  \\
        
        && $\TH_3 \gets \cH(\TH_2 , m_2)$\\
        && $\sk_3 \gets \HKDFExp(\PRK_{3e2m}, 3, \TH_3, \oldnew{\lkey}{\lid})$ \\ 
        && $\oldnew{\IV_3 \gets \HKDFExp(\PRK_{3e2m}, 4,  \TH_3, \lIV)}{\emptyset}$ \\
        && $m_3\gets \oldnew{\cD'(\sk_3, \IV_3; c_3; "")}{\cD(\sk_3, c_3)}$ \\
        && parse $m_3$ as $\oldnew{(\ID_I \| t_3 \| \EAD_3)}{\ID_I}$
        \hl{and $m'_3$ as $(t_3 \| \EAD_3)$} \\
        && $X_s \gets \peerpk[\ID_\sfI]$ \\
        && $\salt_{4e3m} \gets \HKDFExp(\PRK_{3e2m},5, \TH_3, \lH)$ \\
        && $\PRK_{4e3m} \gets \HKDFExt(\salt_{4e3m}, {X_s}^{y_e})$ \\
        && $\accepted \gets 1$ \\
        && $\CTX_3 \gets (\ID_\sfI \| \TH_3 \| X_s \| \EAD_3)$ \\
        && $t'_3 \gets \HKDFExp(\PRK_{4e3m},6, \CTX_3, \oldnew{\lMAC}{\lSEC})$ \\
        && if $t'_3 \neq t_3$ : \textbf{return} $\bot$ \\
 && $\TH_4 \gets \cH(\TH_3, m_3 \mathhl{, m'_3}) $ \\
 && $\sk_4 \gets \HKDFExp(\PRK_{4e3m}, 8, \TH_4, \lkey)$ \\ 
 && $\IV_4 \gets \HKDFExp(\PRK_{4e3m}, 9,  \TH_4, \lIV)$ \\
 && $m_4 \gets ""$, $m'_4 \gets \EAD_4$ \\
 $\TH_4 \gets \cH(\TH_3, m_3 \mathhl{, m'_3})$
 & \sendleft{$c_4, m'_4$}{16mm} & $c_4 \gets \cE'(\sk_4, \IV_4; m_4; m'_4)$ \\
 $\sk_4 \gets \HKDFExp(\PRK_{4e3m}, 8, \TH_4, \lkey)$ \\ 
 $\IV_4 \gets \HKDFExp(\PRK_{4e3m}, 9,  \TH_4, \lIV)$ \\
 if $\cD'(\sk_4, \IV_4; c_4; m'_4) = \bot : \pcreturn \bot$ \\ 

 $\PRK_\out \gets \HKDFExp(\PRK_{4e3m},7, \TH_4, \lH)$ && $\PRK_\out \gets \HKDFExp(\PRK_{4e3m},7, \TH_4, \lH)$ \\
 $\terminated \gets 1$ &&  $\terminated \gets 1$ \\
 $\SK \gets \PRK_\out$ && $\SK \gets \PRK_\out$ \\

    \end{tabular}
    }
    \caption{Optimized \EDHOC with four messages in the \STAT/\STAT Authentication Method. Our modifications compared to  \cite{EDHOC-draft} (draft-ietf-lake-edhoc-15) are represented by \fbox{previous \,|\, \hl{new}} and additions by \hl{gray highlights}} \label{Fig:improvedEDHOC-description}
\end{figure}

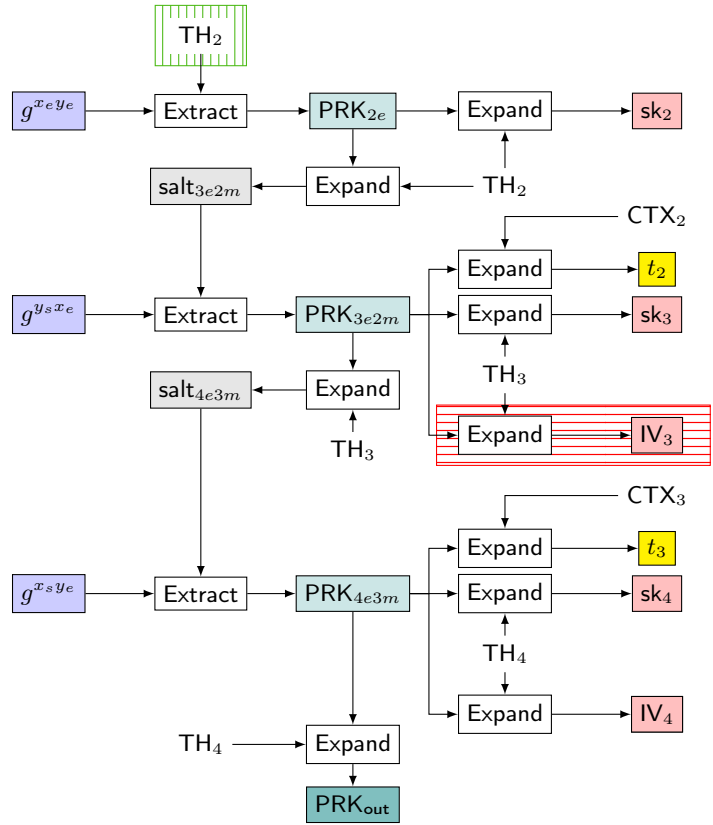
\begin{figure}[h] \centering
    \begin{tikzpicture}
        \definecolor{kellygreen}{rgb}{0.3, 0.73, 0.09}
        \draw[pattern=vertical lines, pattern color=kellygreen, draw=kellygreen] (1.4,8.8) rectangle (2.6,9.6);
        \draw[pattern=horizontal lines, pattern color=red, draw=red] (5.1,3.5) rectangle (8.7,4.3);
        
        \node[draw, fill=blue!20]   at  (0  ,8.2)   (GXY)       {$g^{x_e y_e}$}         ; 
        \node[draw, fill=blue!20]   at  (0  ,5.5)   (GRX)       {$g^{y_s x_e}$}         ; 
        \node[draw, fill=blue!20]   at  (0  ,1.8  )   (GIY)       {$g^{x_s y_e}$}         ;

        \node[draw]                 at  (2  ,8.2)   (XT2)       {\sfExt}                ;
        \node[draw]                 at  (2  ,5.5)   (XT3)       {\sfExt}                ;
        \node[draw]                 at  (2  ,1.8  )   (XT4)       {\sfExt}                ;

        \node[draw, fill=teal!20]   at  (4  ,8.2)   (PRK2)      {$\PRK_{2e}$}           ; 
        \node[draw, fill=teal!20]   at  (4  ,5.5)   (PRK3)      {$\PRK_{3e2m}$}         ; 
        \node[draw, fill=teal!20]   at  (4  ,1.8  )   (PRK4)      {$\PRK_{4e3m}$}         ;
        \node[draw, fill=teal!50]  at  (4  ,-1  )   (PRKout)    {$\PRK_{\out}$}         ;
        
        \node[draw]                 at  (6  ,8.2)   (XP2)       {\sfExp}                ;
        \node[draw]                 at  (6  ,6.1)   (XP3t)       {\sfExp}                ;
        \node[draw]                 at  (6  ,5.5)   (XP3sk)      {\sfExp}                ;
        \node[draw, fill=white]                 at  (6  ,3.9)   (XP3IV3)      {\sfExp}                ;
        \node[draw]                 at  (6  ,2.4)   (XP4t)      {\sfExp}                ;
        \node[draw]                 at  (6  ,1.8)   (XP4sk4)       {\sfExp}                ;
        \node[draw]                 at  (6  ,0.2)   (XP4IV4)       {\sfExp}                ;
        \node[draw]                 at  (4  ,7.2)   (XPs3)      {\sfExp}                ;
        \node[draw]                 at  (4  ,4.5)   (XPs4)       {\sfExp}                ;
        \node[draw]                 at  (4  ,-0.2)   (XPout)      {\sfExp}                ;
        
        \node[draw, fill=gray!20]   at  (2  ,7.2)   (s3e2m)     {$\salt_{3e2m}$}        ;
        \node[draw, fill=gray!20]   at  (2  ,4.5)   (s4e3m)     {$\salt_{4e3m}$}        ;

        \node                       at  (6  ,7.2)   (TH2)       {$\TH_2$}               ;
        \node[fill=white]                       at  (2  ,9.2)   (s2e)       {$\TH_2$}               ;
        \node                       at  (6  ,4.7)   (TH3)       {$\TH_3$}               ;
        \node                       at  (4  ,3.7)   (TH3b)       {$\TH_3$}               ;
        \node                       at  (6  ,1)   (TH4)       {$\TH_4$}               ;
        \node                       at  (2  ,-0.2)   (TH4b)       {$\TH_4$}               ;

        \node                       at  (8  ,6.8)   (ctx2)      {$\textsf{CTX}_2$}  ;
        \node                       at  (8 ,3.1)   (ctx3)      {$\textsf{CTX}_3$}  ;

        \node[draw, fill=pink]      at  (8  ,8.2)   (sk2)       {$\sk_2$}               ;        
        \node[draw, fill=yellow]    at  (8  ,6.1)   (t2)        {$t_2$}                 ;
        \node[draw, fill=pink]      at  (8  ,5.5)   (sk31)       {$\sk_3$}      ;
        \node[draw, fill=pink]      at  (8  ,3.9)   (IV3)       {$\IV_3$}      ;
        \node[draw, fill=pink]      at  (8  ,0.2)   (IV4)       {$\IV_4$}      ;
        \node[draw, fill=yellow]    at  (8  , 2.4)   (t3)        {$t_3$}                 ;
        \node[draw, fill=pink]      at  (8  ,1.8)   (sk4)       {$\sk_4$}      ;
        \node[draw, fill=pink]      at  (8  ,0.2)   (IV4)       {$\IV_4$}      ;
        
        \draw[-latex] (s2e.south) -- (XT2.north) ;
        \draw[-latex] (GXY.east) -- (XT2.west) ; 
        \draw[-latex] (XT2.east)  -- (PRK2.west); 
        \draw[-latex] (PRK2.east) -- (XP2.west) ; 
        \draw[-latex] (XP2.east) -- (sk2.west) ; 
        \draw[-latex] (PRK2.south) -- (XPs3.north) ;
        \draw[-latex] (XPs3.west) -- (s3e2m.east) ; 
        
        \draw[-latex] (s3e2m.south) -- (XT3.north) ;

        \draw[-latex] (GRX.east) -- (XT3.west) ; 
        \draw[-latex] (XT3.east)  -- (PRK3.west); 
        \draw[-latex] (PRK3.east) -- (5,5.5) -- (5,6.1) -- (XP3t.west) ; 
        \draw[-latex] (PRK3.east) -- (XP3sk.west) ; 
        \draw[-latex] (PRK3.east)  -- (5,5.5) -- (5,3.9) --  (XP3IV3.west) ; 
        \draw[-latex] (XP3t.east) -- (t2.west) ; 
        \draw[-latex] (XP3sk.east) -- (sk31.west) ; 
        \draw[-latex] (PRK3.south) -- (XPs4.north) ;
        \draw[-latex] (XPs4.west) -- (s4e3m.east) ; 

        \draw[-latex] (s4e3m.south) -- (XT4.north) ;

        \draw[-latex] (GIY.east) -- (XT4.west) ; 
        \draw[-latex] (XT4.east)  -- (PRK4.west); 
        \draw[-latex] (PRK4.east) -- (5, 1.8) -- (5,2.4) -- (XP4t.west) ; 
        \draw[-latex] (PRK4.east) --  (XP4sk4.west) ; 
        \draw[-latex] (PRK4.east)  -- (5,1.8) -- (5,0.2) --  (XP4IV4.west) ; 
        \draw[-latex] (XP4t.east) -- (t3.west) ; 
        \draw[-latex] (XP4sk4.east) -- (sk4.west) ; 
        \draw[-latex] (XP4IV4.east) -- (IV4.west) ; 
        \draw[-latex] (XP3IV3.east) -- (IV3.west) ; 
        \draw[-latex] (PRK4.south) -- (XPout.north) ;
        \draw[-latex] (XPout.south) -- (PRKout.north) ;
        \draw[-latex] (TH2.north) -- (XP2.south) ; 
        \draw[-latex] (TH2.west) -- (XPs3.east) ; 

        \draw[-latex] (TH3.north) -- (XP3sk.south) ; 
        \draw[-latex] (TH3.south) -- (XP3IV3.north) ; 
        \draw[-latex] (TH3b.north) -- (XPs4.south) ; 

        \draw[-latex] (TH4.north) -- (XP4sk4.south) ; 
        \draw[-latex] (TH4.south) -- (XP4IV4.north) ; 
        \draw[-latex] (TH4b.east) -- (XPout.west);
        \draw[-latex] (ctx2.west) -- (6,6.8) -- (XP3t.north) ; 
        \draw[-latex] (ctx3.west) -- (6,3.1) -- (XP4t.north) ; 
        
    \end{tikzpicture}
    \caption{Key Derivation (for the \STAT-\STAT Method) from~\cite{EDHOCFormal21}. Green vertical hatchs denote additions and red horizontal hatchs denote removals compared to the initial version. \label{Fig:Key-Schedule-Improved}}
\end{figure}

We here make some remarks on the initial protocol, with some improvements that appear in gray highlights in Figure~\ref{Fig:improvedEDHOC-description}, and to the removed/additional hatched patterns in Figure~\ref{Fig:Key-Schedule-Improved}.

\subsection{On Mutual Authentication}
The encryption key $\sk_3$, used by the initiator to encrypt its second message $m_3$, is computed by calling \HKDFExp on $\PRK_{3e2m}$. 
However, even an adversary that plays in the name of a non-corrupted user, is able to compute $\PRK_{3e2m}$, when knowing the Initiator ephemeral key $x_e$, as $\PRK_{3e2m}$ does not depend on $x_s$, the long term secret key of the Initiator. In order to break the Initiator authentication, with respect to a Responder, an adversary can play on behalf of any user as an Initiator. It will be able to compute $\sk_3$, but not $t_3$, for which value it will need some luck, but this is only 64-bit long! Which is not enough for a 128-bit security.

To get around this issue, we suggest to modify the construction of Initiator's second message as follows:
Initial message $m_3 = (\ID_\sfI \| t_3  || \EAD_3)$ is split as $m_3 \gets (\ID_\sfI)$ and $m'_3 \gets (t_3  || \EAD_3)$. 
Thus, $m_3$ is encrypted using $\sk_3$ (with a one-time pad encryption scheme $\varPi=(\cE,\cD)$, under $\sk_3$ still depending on $\PRK_{3e2m}$) into $c_{3}$. Then $m'_3$ does not need to be encrypted.  We introduce the value $\lSEC$, always set as the expected bit-security parameter, independently of the $\lMAC$ value. Then, we set the length of $t_3$ to be $\lSEC$, as it already authenticates $\CTX_3 = (\ID_\sfI \| \TH_3 \| X_s \| \EAD_3)$.
Concretely, the second message sent by the initiator to the responder is:

$$c_3 \| m'_3\text{, where }c_3 = \cE(\sk_3, m_3), m'_3 = t_3  || \EAD_3$$

Once the Responder receives $(c_3, m'_3)$, he first decrypts $c_{3}$, retrieves $X_s$ using $m_3$, computes $\PRK_{4e3m}$ and is then able to verify the tag $t_3$, allowing to check the authenticity of $\ID_\sfI$, as well as all the other values is $\CTX_3 = \ID_\sfI \| \TH_3 \| X_s \| \EAD_3$.
The extra required length for the tag $t_3$ is perfectly compensated by the absence of the tag jointly sent when using Authenticated Encryption, and the plaintext length of $m_3$ is the same as the encryption of $m_3$. Therefore, this does not impact the communication cost of the protocol, until $\lSEC \leq 2 \times \lMAC$, but improves to $\lSEC$-bit security for Initiator-Authentication.

About the Responder-Authentication, $t_2$ also provides a 64-bit security level only: by guessing it, any active adversary can make the initiator terminate, and thus breaking the responder-authentication, if one does not wait for the fourth flow $c_4,m'_4$. However, with this fourth flow, we can show the $2 \times \lMAC$-bit security level is achieved.

\colorlet{hlgray}{gray!20}
\sethlcolor{hlgray}

\subsection{On Reduction Efficiency}
After analysis, we also notice another improvement:
the key $\PRK_{2e}$ is computed according to $g^{x_e y_e}$ only, as the salt used in \HKDFExt is an empty string. 
When considering several parellels sessions, this allows an adversary to find a collision with any of the session making a single call to \HKDFExt. 
Therefore, we replace the empty string used as salt with $\TH_2$ that depends on the session variables and is different for each session. 
Thus, an adversary has to make a call to $\HKDFExt$ with a chosen $\TH_2$, linked to a specific session. This makes the reduction cost of the key-privacy game independent of $n_\sigma$, the number of sessions.

\bibliographystyle{alpha}
\bibliography{biblio, cryptobib/abbrev3, cryptobib/crypto}

\end{document}